\documentclass{aa}
\usepackage{multirow}
\usepackage{txfonts}
\usepackage{graphicx}
\usepackage{xcolor}

\def\H2{H$_2$}

\begin{document}

\title{Glycolaldehyde in Perseus young solar analogs\thanks{Based on observations carried out with the IRAM Plateau de Bure interferometer. IRAM is supported by INSU/CNRS (France), MPG (Germany), and IGN (Spain)}}
\author{M. De Simone\inst{1} \and  
C. Codella \inst{1} \and 
L. Testi \inst{1,2,3,4} \and
A. Belloche \inst{5} \and
A.J. Maury \inst{6} \and 
S. Anderl  \inst{7,8} \and
Ph. Andr\'e \inst{6} \and
S. Maret \inst{7,8} \and 
L. Podio \inst{1}}

\institute{
INAF, Osservatorio Astrofisico di Arcetri, Largo E. Fermi 5,
50125 Firenze, Italy
\and
ESO, Karl Schwarzchild Str. 2, 85748 Garching bei M\"unchen, Germany
\and   
Excellence Cluster `Universe', Boltzmannstr. 2, D-85748 Garching bei Muenchen, Germany
\and 
  {Gothenburg Center for Advance Studies in Science and Technology,
Department of Mathematical Sciences, Chalmers University of Technology and University of Gothenburg, SE-412 96 Gothenburg, Sweden}
\and
Max-Planck-Institut f\"ur Radioastronomie, Auf dem H\"ugel 69, 53121 Bonn, Germany
\and
Laboratoire AIM-Paris-Saclay,
CEA/DSM/Irfu - CNRS -Universit\'e
Paris Diderot, CE Saclay, 91191 Gif-sur-Yvette Cedex, France
\and
Univ. Grenoble Alpes, CNRS, IPAG, 38000, Grenoble, France 
\and
CNRS, IPAG, 38000, Grenoble, France 
}

\offprints{C. Codella, \email{codella@arcetri.astro.it}}
\date{Received date; accepted date}

\authorrunning{De Simone et al.}
\titlerunning{Glycolaldehyde in Perseus young solar analogs}

\abstract
{The earliest evolutionary stages of low-mass protostars are characterised by the so-called
hot-corino stage, when the newly born star heats the surrounding material and enrich 
the gas chemically. Studying this evolutionary phase of solar protostars may help understand the evolution of prebiotic complex molecules in the development of planetary systems.}
{In this paper we focus on the occurrence of glycolaldehyde (HCOCH$_2$OH) in young solar analogs by performing the first homogeneous and unbiased study of this molecule in the Class 0 protostars of the nearby Perseus star forming region.}
{We obtained sub-arcsec angular resolution maps at 1.3mm and 1.4mm of   
glycolaldehyde emission lines using the IRAM Plateau de Bure (PdB) interferometer in the framework 
of the CALYPSO IRAM large program.}
{Glycolaldehyde has been detected towards 3 Class 0 and 1 Class I protostars out
of the 13 continuum sources targeted in Perseus: NGC1333-IRAS2A1, NGC1333-IRAS4A2,
NGC1333-IRAS4B1, and SVS13-A. The NGC1333 star forming region looks 
particularly glycolaldehyde rich, with a rate of occurrence up to
60\%.
The glycolaldehyde spatial distribution overlaps with the continuum one, tracing the inner 
100 au around the protostar. A large number of lines (up to 18), 
with upper-level energies  $E_{\rm u}$ from 37 K up to 375 K has been detected.
We derived column densities $\geq$ 10$^{15}$ cm$^{-2}$ and rotational temperatures $T_{\rm rot}$ between 115 K and 236 K, imaging for the first time hot-corinos around NGC1333-IRAS4B1 and SVS13-A.}
{In multiple systems glycolaldehyde emission is detected only in one component.
The case of the SVS13-A+B and IRAS4-A1+A2 systems support that the detection of glycolaldehyde (at least in the present Perseus sample) indicates older protostars (i.e. SVS13-A and IRAS4-A2),  
evolved enough to develop the hot-corino region (i.e. 100~K in the inner 100~au).
However, only two systems do not allow us to firmly conclude whether the 
primary factor leading to the detection of 
glycolaldehyde emission is the environments hosting the protostars, evolution (e.g. low value of $L_{\rm submm}$/$L_{\rm int}$), or accretion 
luminosity (high $L_{\rm int}$).}


\maketitle

\section{Introduction}

The gas surrounding a new born star can reach temperatures
higher than 100 K, due to thermal heating by the protostar
(the so called hot-corinos see e.g. Ceccarelli et al. 2007)
as well as shocks produced by both accretion and ejection processes
(e.g. Codella et al. 2009; Sakai et al. 2014).
In turn, such temperatures favour a rich complex
organic chemistry, with a consequent dramatic
increase of the abundance of the so-called Complex Organic Molecules 
(COMs; i.e. organic molecules with at least 6 atoms detected in space,
Herbst \& van Dishoeck 2009), 
which can be considered the first bricks
to build biologically relevant molecules.
Among the COMs, glycolaldehyde (HCOCH$_2$OH)
is the simplest sugar-like molecule and it is expected to be involved in the chemical
processes leading to ribose 
(e.g. Weber 1998; Jalbout et al. 2007), 
in turn the backbone of terrestrial RNA.

The first detection of interstellar glycolaldehyde was made in the
hot core Sagittarius B2(N) located in the
high-mass star forming region Sagittarius B2 close to
the Galactic Centre (Hollis et al. 2000, 2004; Halfen et al. 2006;
Requena-Torres et al. 2008). 
Later on, Beltr\'an et al. (2009) 
imaged glycolaldehyde emission for the first time towards
a star forming region, namely the massive hot core
(the scaled-up version of a hot-corino) G31.41+0.31,
while Calcutt et al. (2014) reported glycolaldehyde
towards further hot cores.
However, the large distance of the observed high-mass
star forming regions ($\geq$ 1 kpc) hampers
(i) to assess a reliable association with one or more
young stellar objects due to multiplicity, and (ii) an analysis
on spatial scales comparable to that of a protoplanetary disk
(i.e. $\sim$ 100 au). In addition, the study of the formation and 
evolution of complex organic molecules (including glycolaldehyde) in young 
Solar analogs is essential to understand the potential of formation and
evolution of pre-biotic material in these systems.

J\o{}rgensen et al. (2026) reported the first detection of
glycolaldehyde emission with ALMA towards the young Solar analog 
low-mass Class 0 protostellar system 
IRAS 16293-2422.  
They detected 13 emission lines indicating the presence of glycolaldehyde in the
warm gas (200--300 K) in the inner regions of the system, within $\sim 50$~au from each 
protostar. This study paved the way to the study
of glycolaldehyde around Sun-like star forming 
regions (see also the recent work by J\o{}rgensen et al. 2016).
Indeed, Coutens et al. (2015) 
and Taquet et al. (2015) 
detected, using the IRAM PdB array,
up to 8 glycolaldehyde lines towards two others well know hot-corinos associated with the
NCC1333-IRAS2A and NGC1333-IRAS4A protostars, finding temperatures higher than 100 K.
Going beyond these initial findings, it is important to 
perform a more systematic study to understand
how  common is the occurrence of glycolaldehyde around solar-type protostars,
and whether the presence of relatively
abundant gas-phase glycolaldehyde can be linked to a specific 
evolutionary phase. A uniform sensitivity and high angular resolution 
survey of protostars is the best approach to attempt to answer these 
questions.

The CALYPSO\footnote{http://irfu.cea.fr/Projects/Calypso} 
survey of Class 0 protostars 
(i.e. 10$^4$--10$^5$~yr solar analog protostars; Andr\'e et al. 1993, 2000), 
which was carried out with the IRAM-PdB interferometer 
(see Maury et al. 2014, for more details), offers a unique dataset 
for studying glycolaldehyde.
The sources are mainly Class 0 protostars with the exception of SVS13-A which
has apparently entered in the Class I stage (Lada 1987).  
CALYPSO includes all Solar-mass protostellar systems in the
Perseus L1448 
($d$ = 232$\pm$18 pc; Hirota et al. 2011) 
and NGC-1333 
($d$ = 235$\pm$18 pc; Hirota et al. 2008) regions.
The dataset (see Sect.~2) includes high 
angular resolution ($\leq$ 1$\arcsec$) spectral maps at 1.3 and 1.4~mm,
which cover a broad range of glycolaldehyde lines and upper-level energy and are adequate for 
an initial statistical study.
In this paper we present an analysis of these data, limited to the 
glycolaldehyde molecule, a detailed analysis of the continuum data will be
presented in Maury et al.~(in preparation), and the analysis of all complex organic
molecules detected in the survey in Belloche et al.~(in preparation).

The goal of the present paper is: 
(i) to perform a first statistical analysis of the occurrence of glycolaldehyde emission towards Class 0 protostellar sources, and
(ii) to take advantage of the combination of high sensitivity and high spatial resolution 
provided by the CALYPSO database to detect 
a large number of lines covering a wide range of upper-level energy, leading to reliable estimates of the rotational (excitation) temperatures and glycolaldehyde column densities.
The observations are reported in Sect. 2, 
while in Sect. 3 we describe the selected sample. The obtained images, the derived physical conditions, and the estimate of the glycolaldehyde abundances are reported in Sect. 4. 
Our conclusions are summarized in Sect. 5. 

\section{Observations} 
\label{sec:obs}

The Perseus sources (listed in Table 1) were observed at 1.3mm and 1.4mm
with the IRAM PdB six-element array
during several tracks between November 2010 and February 2012 using both the A and C configurations.
The shortest and longest baselines are 19 m and 762 m, respectively, allowing us to recover
emission at scales from $\sim$ 8$\arcsec$ down to $\sim$ 0$\farcs$4.
The glycolaldehyde lines\footnote{Spectroscopic parameters 
have been extracted from the Jet
Propulsion Laboratory molecular database (Pickett et al. 1998), 
see Tables 2 to 5.}
were observed using the WideX backend to cover two 4-GHz spectral windows
(one at 1.3mm and one at 1.4mm) with a spectral resolution of 
2 MHz ($\sim$ 2.6 km s$^{-1}$ at 1.4 mm).
The observed spectral ranges are the following:
217.0--220.5 GHz (1.4mm hereafter), and 229.0--233.0 GHz (1.3mm).
Calibration was carried out following standard procedures,
using GILDAS-CLIC\footnote{http://www.iram.fr/IRAMFR/GILDAS}.
The phase rms was $\le$ 65$\degr$,
the precipitable water vapor (PWV) was typically less than 2mm,
the system temperatures less than 200 K.
The final uncertainty on the absolute flux scale is $\leq$ 15\%.
The continuum emission was removed from the visibility tables
to produce continuum-free line tables.
The typical rms noise in the 2-MHz channels was 3--9 mJy beam$^{-1}$.
Images were produced using natural weighting, and restored with a 
clean beam of $\leq$ 1$\arcsec$, reported, for each source and for 
each wavelength, in Table 1.

\section{The sample}

\begin{table*}
\caption{Continuum peak emission at 1.4mm, detected in Perseus using the CALYPSO survey, 
where glycolaldehyde emission has been searched for.}
\centering
\begin{tabular}{lcccccccc}
\hline
Source	& $\alpha({\rm J2000})$$^a$  & $\delta({\rm J2000})$$^a$ & $V_{\rm sys}$$^a$ & $d$$^b$ & $F_{\rm 1.4mm}$ & $L_{int}$$^c$ & \multicolumn{2}{c}{$HPBW_{\rm syn}$}  \\
		& ($^h$ $^m$ $^s$) & ($\degr$ $\arcmin$ $\arcsec$) & (km s$^{-1}$) & (pc) & (mJy/beam) &  ($L_{\odot}$) &  \multicolumn{2}{c}{$\arcsec$$\times$$\arcsec ($\degr)} \\
\cline{8-9}
 & & & & & & & 1.4mm & 1.3mm \\
\hline
L1448-2A & 03:25:22.406 & +30 45 13.28 & +4.2 & 232 & 31 & 2.7 & 0$\farcs$99$\times$0$\farcs$71 (28$\degr$) & 1$\farcs$03$\times$0$\farcs$81 (39$\degr$) \\
L1448-NB & 03:25:36.364 & +30 45 14.84 & +4.7 & 232 & 25 & 4.1 & 0$\farcs$81$\times$0$\farcs$67 (67$\degr$) & 0$\farcs$65$\times$0$\farcs$48 (48$\degr$) \\
L1448-NA & 03 25 36.503 & +30 45 21.87 & +4.7 & 232 & 139 &  - & 0$\farcs$81$\times$0$\farcs$67 (67$\degr$) & 0$\farcs$65$\times$0$\farcs$48 (48$\degr$) \\
L1448-C & 03 25 38.878 & +30 44 05.32 & +5.2 & 232 & 102 & 6.8 & 0$\farcs$77$\times$0$\farcs$63 (22$\degr$) & 0$\farcs$61$\times$0$\farcs$38 (33$\degr$) \\
IRAS2A3 & 03 28 55.514 & +31 14 34.84 & +7.3 & 235 & 16 &  - & 0$\farcs$82$\times$0$\farcs$80 (32$\degr$) & 0$\farcs$75$\times$0$\farcs$74 (168$\degr$) \\
IRAS2A1 & 03 28 55.575 & +31 14 37.05 & +7.3 & 235 & 99 & 26.7 & 0$\farcs$82$\times$0$\farcs$80 (32$\degr$) & 0$\farcs$75$\times$0$\farcs$74 (168$\degr$) \\
IRAS2A2 & 03 28 55.677 & +31 14 35.56 & +7.3 & 235 & 12 & - & 0$\farcs$82$\times$0$\farcs$80 (32$\degr$) & 0$\farcs$75$\times$0$\farcs$74 (168$\degr$) \\
SVS13-B & 03 29 03.075  & +31 15 51.71 & +8.4 & 235 & 122 & 1--2$^c$ & 1$\farcs$06$\times$0$\farcs$80 (20$\degr$) & 0$\farcs$70$\times$0$\farcs$42 (20$\degr$) \\
SVS13-A & 03 29 03.759  & +31 16 03.74 & +8.4 & 235 & 123 & 24.5 & 1$\farcs$06$\times$0$\farcs$80 (20$\degr$) & 0$\farcs$70$\times$0$\farcs$42 (20$\degr$) \\
IRAS4A2 & 03 29 10.429  & +31 13 32.10 & +7.2 & 235 & 284 & - & 1$\farcs$10$\times$0$\farcs$81 (20$\degr$) & 0$\farcs$70$\times$0$\farcs$42 (20$\degr$) \\
IRAS4A1 & 03 29 10.531  & +31 13 30.96 & +7.2 & 235 & 795 & 2.8 & 1$\farcs$10$\times$0$\farcs$81 (20$\degr$) & 0$\farcs$70$\times$0$\farcs$42 (20$\degr$) \\
IRAS4B1 & 03 29 12.012  & +31 13 08.07 & +6.7 & 235 & 449 & 1.0 & 1$\farcs$08$\times$0$\farcs$83 (20$\degr$) & 0$\farcs$70$\times$0$\farcs$42 (20$\degr$) \\
IRAS4B2 & 03 29 12.840  & +31 13 06.98 & +6.7 & 235 & 53 & - & 1$\farcs$08$\times$0$\farcs$83 (20$\degr$) & 0$\farcs$70$\times$0$\farcs$42 (20$\degr$) \\
\hline
\end{tabular}

$^a$ Positions of the 1.3mm continuum peak emission and systemic velocities are extracted 
from the CALYPSO dataset (Codella et al. 2014; Santangelo et al. 2015; Anderl et al. 2016; 
Maury et al., in preparation). 
$^b$ From Hirota et al. (2008, 2011).
$^c$ Internal luminosities were derived from the observed flux at 70 $\mu$m 
(Dunham et al. 2008), a wavelength at which {\it Herschel} Gould Belt survey observations 
(Andr\'e et al. 2010) provide data at 8$\arcsec$ spatial resolution
(see also Ladjelate et al., in preparation).
The internal luminosity of SVS13-B is more uncertain
due to the proximity to SVS13-A.
\end{table*}

The sample of sources is based on all the peaks detected in the 1.3mm and 1.4mm 
continuum images of CALYPSO targets in Perseus (see Codella et al. 2014; Santangelo et al. 2015; Anderl et al. 2016; Maury et al., in preparation). We selected the sources in the Perseus complex, which is a
classical laboratory to study low-mass star formation; in particular, the selected
sources belong to the most active sites of current star formation in the Perseus cloud: the NGC-1333 and L1448 clusters. We observed 4 multiple systems in NGC-1333 (IRAS2A, IRAS4A, IRAS4B, and SVS13) as well as 4 protostars in L1448 (2A, NB, NA, and C objects), for a total number of 13 objects to be analysed to search for glycolaldehyde emission. 
In Table 1 we report the continuum peaks detected at 1.4 mm in Perseus using the CALYPSO database; for every source are reported the positions of the peaks and the systemic velocities (V$_{sys}$) extracted from the CALYPSO dataset, the 1.4 mm peak flux, the internal luminosity ($L_{\rm int}$), and the clean beams (HBPW$_{syn}$) for each wavelength. The internal luminosities are derived using the 70 $\mu$m measurements provided by the {\it Herschel} Gould Belt survey (Andr\'e et al. 2010).
In particular, Dunham et al. (2008), in the context of the Spitzer Space Telescope
Legacy Project "From Molecular
Cores to Planet Forming Disks", showed how the 70 $\mu$m
flux and internal luminosity of a protostar are
tightly correlated. As the former is a directly observable quantity
but the latter is not, this correlation gives a powerful method for
estimating protostellar internal luminosities. 
$L_{\rm int}$ is expected to be a more reliable probe of accretion luminosity than the bolometric luminosity $L_{\rm bol}$ (see also Ladjelate et al., in preparation), which requires a full coverage of the Spectral Energy Distribution (SED) at high-spatial resolution (8$\arcsec$).

The continuum map of NGC1333-IRAS2A reveals the main protostar IRAS2-A1, plus two weaker peaks, IRAS2-A2 and IRAS2-A3 (see Codella et al. 2014). The nature of A2 and A3 is still debated, being possibly dust fragments associated with 
the molecular cloud and swept up by the outflows driven by the nearby IRAS2-A1 protostar
(see Codella et al. 2014; Tobin et al. 2016). For the purpose of the analysis presented in this paper, we have considered these as genuine protostars, except where explicitly mentioned otherwise.
The NGC-1333-IRAS4 system has two main objects: IRAS4A and IRAS4B.
In turn, IRAS4A is a binary composed by two Class 0 objects,
separated by 1$\farcs$8: IRAS4-A1 and IRAS4-A2
(e.g. Looney et al. 2000).
While IRAS4-A1 is more than three times brighter in the mm-flux than its companion, only IRAS4-A2 shows a high molecular complexity (see Taquet et al. 2015; Coutens et al. 2015; Santangelo et al. 2015). IRAS4B is located $\sim$ 30$\arcsec$ southeast of IRAS4A, and it is associated with two compact continuum sources, B1 and B2, with an angular separation of 11$\arcsec$ 
(e.g. Looney et al. 2000; Choi et al. 2011).
NGC1333-SVS13 is a cluster dominated at millimeter wavelengths by two sources: SVS13-A, 
and SVS13-B. SVS13A is a protostar already in the Class I stage and drives 
an extended outflow associated with the well-known HH7-11 chain
(e.g. Lefloch et al. 1998; Chen et al. 2009).
On the other hand, SVS13-B is an earlier protostar lying  southwest of SVS13-A 
at a distance of 15$\arcsec$
(e.g. Bachiller et al. 1998; Looney et al. 2000). Finally, the L1448-NA, L1448-NB, and L1448-2A protostars are located in the northern portion of the L1448 complex, at about $\sim$ 1$\degr$ southwest of NGC1333,
while L1448-C is located at the center of the L1148 complex (see Looney et al. 2000; Tobin et al. 2016, and references therein).

\section{Results and discussion}

\begin{figure*}
\begin{center}
\includegraphics[scale=0.32]{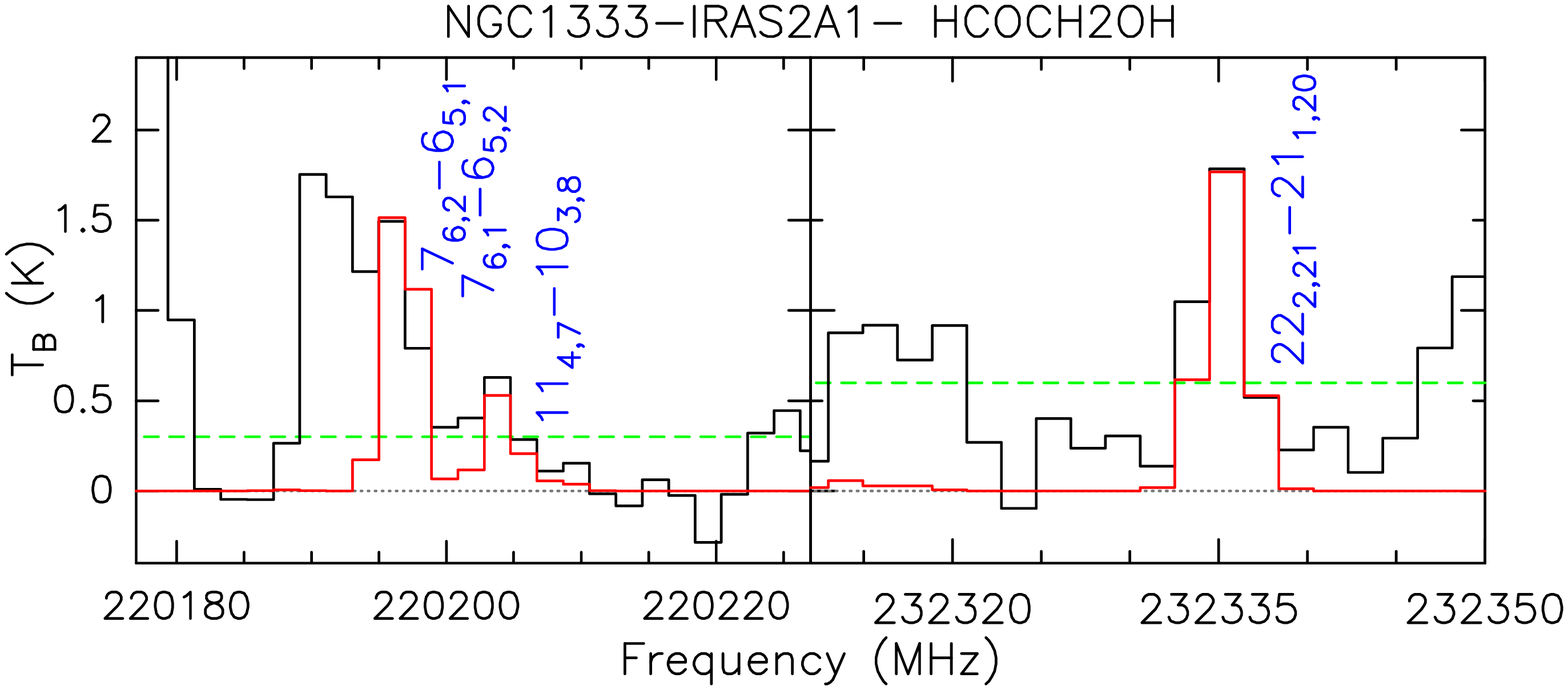} 
\includegraphics[scale=0.32]{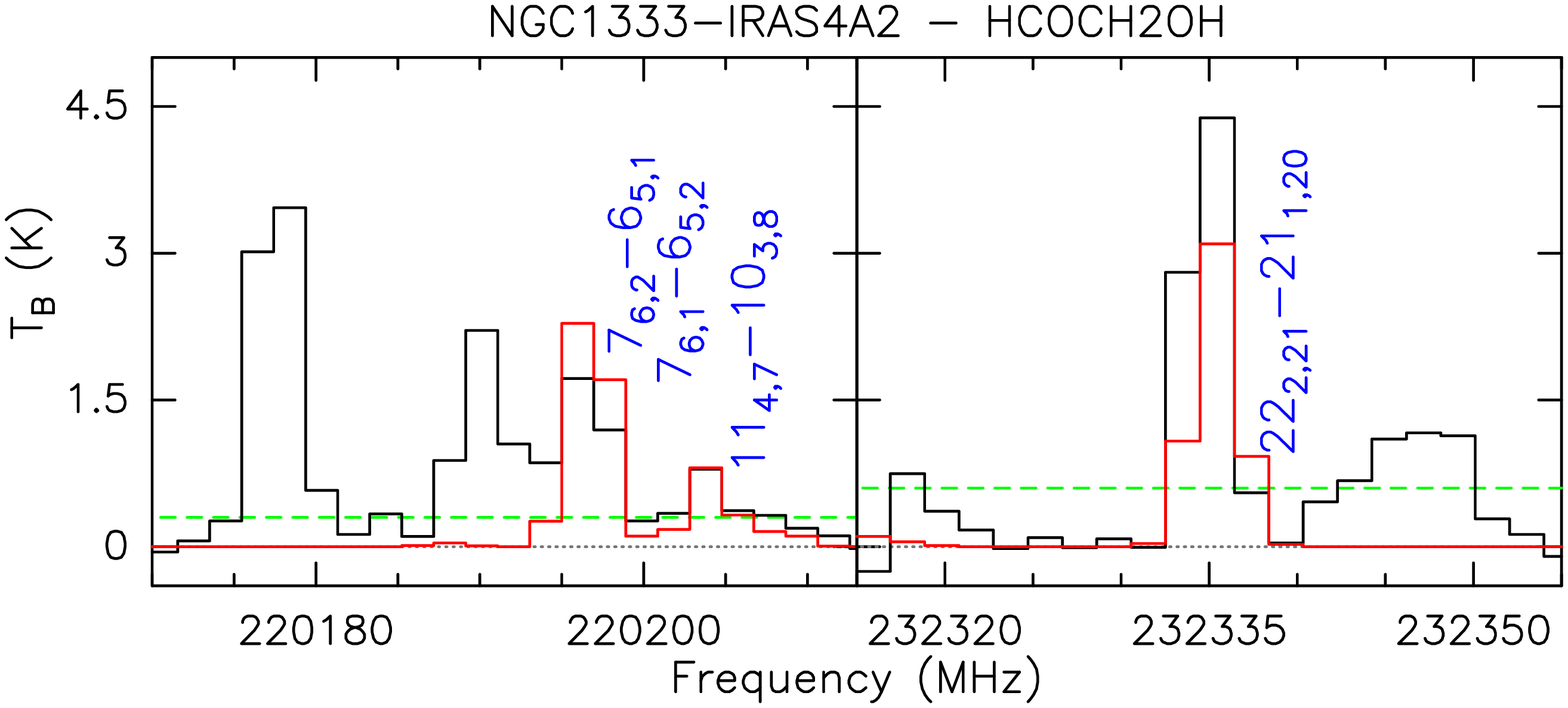} \\ 
\includegraphics[scale=0.32]{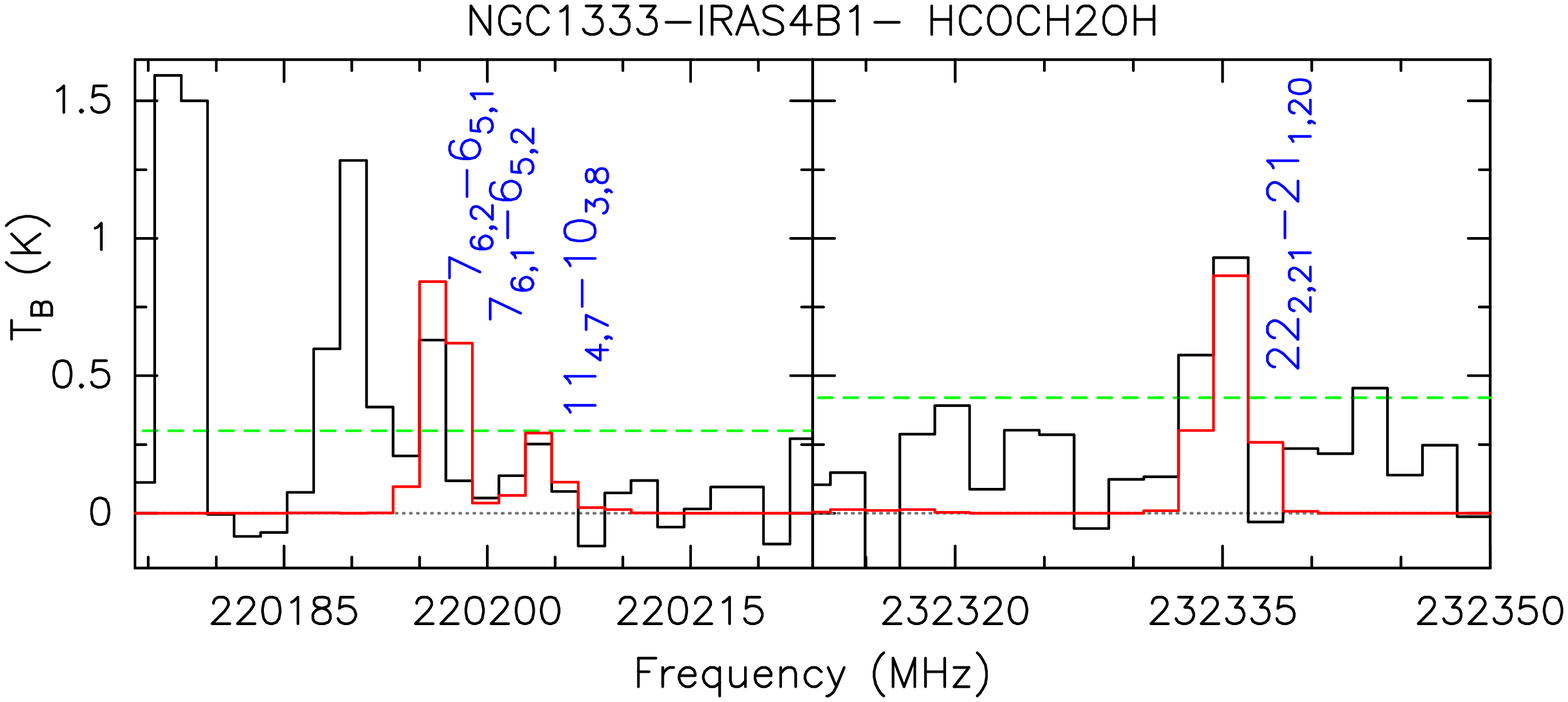} 
\includegraphics[scale=0.32]{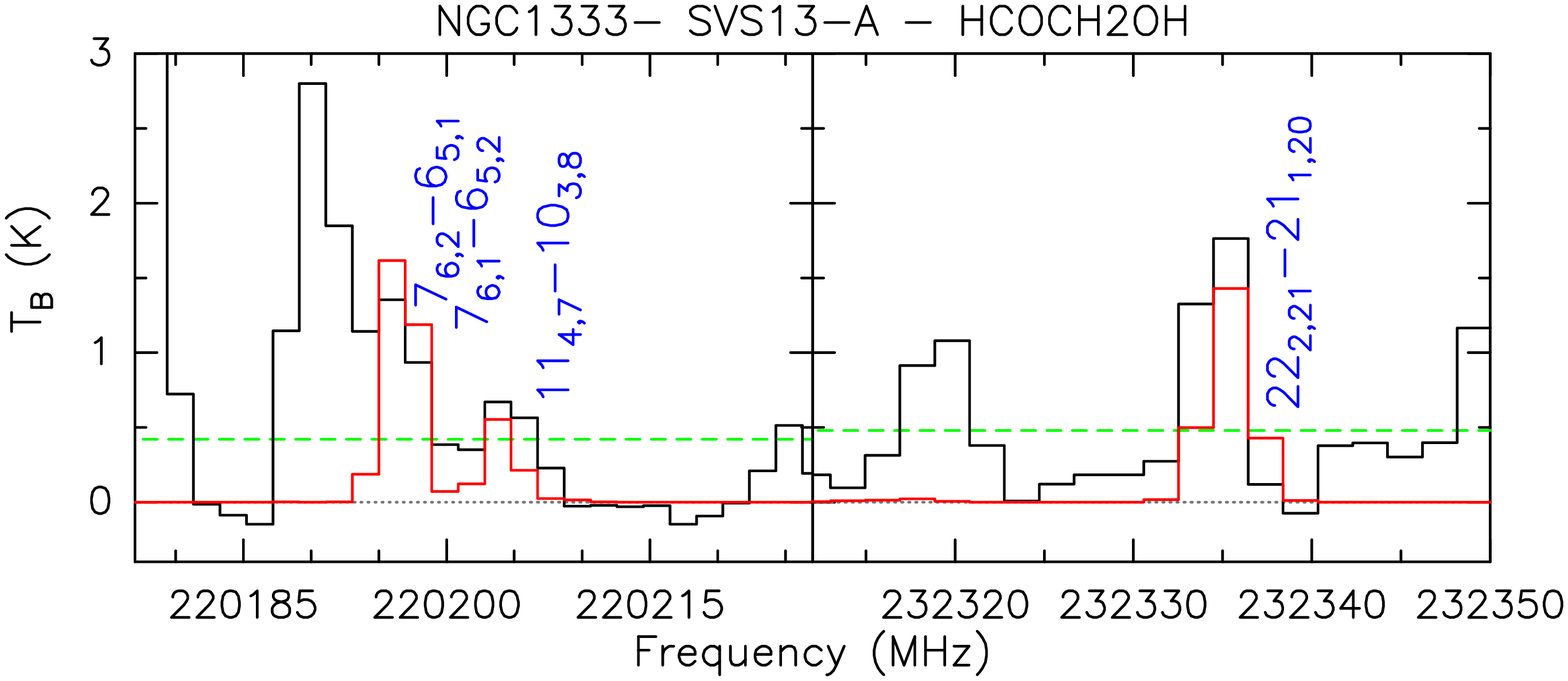}
\caption{Examples of glycolaldehyde emission lines (in $T_{\rm B}$ scale)
observed in the 1.3mm and 1.4mm spectral windows 
(and at different upper-level energy, $E_{\rm u}$ from 37 K to 135 K) towards 
NGC1333-IRAS2A1 ({\it Upper-left panel}), NGC1333-IRAS4A2
({\it Upper-right panel}), NGC1333-IRAS4B1 ({\it Lower-left panel}),
and NGC1333-SVS13A ({\it Lower-right panel}). 
The horizontal green dotted lines show the 3$\sigma$ level.
In blue we mark the glycolaldehyde lines extracted from Tables A1-A4.
The bright line at $\sim$ 220190 MHz is 
due to CH$_3$OCHO(17$_{\rm 4,13}$--16$_{\rm 4,12}$)A 
(Maury et al. 2014; Belloche et al. in preparation).
The red line shows the synthetic spectrum obtained with the GILDAS--Weeds package (Maret et al. 2011) and assuming the rotation diagram solutions (see Fig. 3).}
\end{center}
\end{figure*}

The results of the present search towards the coordinates listed in Table 1 
indicates the presence of
glycolaldehyde towards 4 out of the 13 observed continuum peaks.
If we consider that the 
classification of the IRAS2-A3 and IRAS2-A2 continuum peaks as genuine protostars is debated 
(see the recent paper by Tobin et al. 2016), the detection
rate is $\sim$36\%. 
Interestingly, glycolaldehyde has only been detected 
towards the NGC1333 sources, where the detection rate is as high as $\sim$60\%.

\begin{table}
\caption{Number of glycolaldehyde lines and results
of the LTE analysis for the detected sources.}

\centering
\begin{tabular}{lcccccc}
\hline
Source	& $N_{\rm trans}$ & $E_{\rm u}$ & $T_{\rm rot}$ & $N_{\rm tot}$ \\
		&  & (K) & (K) & (10$^{14}$ cm$^{-2}$)\\
\hline
IRAS2A1 & 13 & 37--318 & 159(24) & 46(20) \\
IRAS4A2 & 18 & 37--375 & 236(74) & 88(70) \\
IRAS4B1 & 13 & 37--271 & 152(35) & 15(9) \\
SVS13-A & 11 & 37--375 & 115(13) & 21(8)  \\
\hline	
\end{tabular}

$^a$ In several cases (3 for IRAS2A1, 1 for IRAS4A2, IRAS4B1, and SVS13-A; see Tables A1 to A4), the detected lines are due to two glycolaldehyde transitions at frequencies
blended at the present spectral resolution.
\end{table}

Figures A.1 to A.9 report the spectra at the frequencies of the 
glycolaldehyde lines observed towards IRAS2A1, SVS13-A, IRAS4A2, and IRAS4B1.
The spectral resolution does not allow us to probe the profiles of the emission
lines and the observed spectra are affected by line blending.
Nevertheless, there is little doubt that the large WideX bandwidths 
(8 GHz in total) allowed us to detect towards the four sources a considerable number of glycolaldehyde lines
(up to 18) with a signal-to-noise ratio (S/N) of at least 3.
We then selected only those transitions with a corresponding synthetic line (derived using
the rotation diagram solutions, see below) that reproduces at least 
50\% of a detected peak temperature. 
The lines were identified using the Jet
Propulsion Laboratory (JPL) molecular database 
(Marstokk \& M{\o}llendal 1970, 1973; Pickett et al. 1998; Butler et al. 2001; Widicus Weaver et al. 2005; Carroll et
al. 2010).
Figure 1 shows examples of glycolaldehyde emission lines (in $T_{\rm B}$ scale) observed in the 1.3mm and 1.4mm spectral windows towards the four detected sources. 
Figures A.1, A.3, A.5, and A.7 show all the lines used for the following analysis.
Tables A.1-A.4 report the observed line parameters; we list both peak line intensities ($T_{\rm peak}$) and velocities ($V_{peak}$, consistent with
the LSR velocities of the objects), linewidth ($FWHM$), and the integrated emission ($\int$ $T dv$). All the values are in $T_{\rm B}$ scale and are derived from Gaussian fits.
A wide upper-level energy range is observed (see Table 2), with $E_{\rm u}$ from 37 K up to 375 K: this allows us 
(i) to improve on the results of 
Coutens et al. (2015) and Taquet et al. (2015) 
on IRAS2A1 and IRAS4A2, by increasing the number of detected lines
and expanding the range of upper-level energy probed, 
and
(ii) to firmly assess, for the first time, 
the glycolaldehyde occurrence towards IRAS4B1 and SVS13-A.

\subsection{HCOCH$_{2}$OH spatial distributions}

Figure 2 compares the spatial distribution of the
glycolaldehyde emission (colour scale and green contours) with that of the 1.4mm continuum (white contours; see Codella et al. 2014;
Maury et al., in preparation) 
towards (from top to bottom) NGC1333-IRAS2A1, NGC1333-IRAS4A, NGC1333-IRAS4B1, and NGC1333 SVS13-A. 
The glycolaldehyde distribution refers to emission lines 
in both the 1.3mm and 1.4mm spectral windows, i.e. 
the sum of the 7$_{6,2}$-6$_{5,1}$ and 7$_{6,1}$-6$_{5,2}$ lines (220.2 GHz, $E_{\rm u}$ = 37 K), and the
 22$_{2,21}$-21$_{1,20}$ line (232.3 GHz, $E_{\rm u}$ = 135 K). 
Consistent with the results of Taquet et al. (2015), our images
indicate that the glycolaldehyde emission is 
only marginally resolved. 
For each source, we derived the emission size of the brightest line at
1.3 mm shown in Fig. 2 by fitting an elliptical Gaussian in the $uv$ domain: we
obtain about 
0$\farcs$5 (IRAS2A1), 0$\farcs$4 (IRAS4A2, IRAS4B1), and 0$\farcs$3 (SVS13-A), which implies
that the glycolaldehyde emission is confined in the inner 100~au around each
protostar. 
The glycolaldehyde emission peak is consistent, considered the angular resolution, with the continuum peaks. Our results confirm
the hot-corino nature of NGC1333-IRAS2A1 and NGC1333-IRAS4A2, previously
noted by Maury et al. (2014) and Taquet et al. (2015), and
in addition supports the same nature for IRAS4B1 and SVS13-A 
(see Sect. 4.2 for the temperature measurements).

Finally, the image of the 
brightest glycolaldehyde emission (at 220.2 GHz) of the 
IRAS4-A1+A2 system, shows, in addition to the A2 compact
core, a tentative detection at 5$\sigma$ of a 
more extended emission. If significant, 
either this could trace part of the extended envelope 
hosting the two protostars; 
or it could be associated with the swept-up material associated with the
two N-S outflows driven by the A1 and A2 protostars
(Santangelo et al. 2015). In the latter case, this would be the
first signature of glycolaldehyde emission from a protostellar outflow.
However, note that this feature could be potentially 
contaminated by emission due to the SO$^{17}$O(13$_{\rm 2,12}$--13$_{\rm 1,13}$) transition at 220196.75 MHz ($E_{\rm u}$ = 93 K, log ($A_{\rm ij}$/s$^{-1}$) = -3.8). We inspected the CALYPSO dataset searching for emission lines due to SO$_2$ isotopologues.  We have only a weak (less than 3$\sigma$) $^{34}$SO$_2$(4$_{\rm 2,2}$--3$_{\rm 1,3}$) emission  
at 229.9 GHz ($E_{\rm u}$ = 19 K, log ($A_{\rm ij}$/s$^{-1}$) = -4.1), but only around the NGC1333-IRAS4A2
protostar.
In addition, again only IRAS4A2, we have 
a tentative ($\geq$ 3$sigma$) detection
of the SO$_2$(11$_{\rm 5,7}$--12$_{\rm 4,8}$) line ($E_{\rm u}$ = 122 K) at 229.3 GHz, and, not surprisingly, no detection of the SO$_2$(22$_{\rm 7,15}$--23$_{\rm 6,18}$) line at 219.3 GHz ($E_{\rm u}$ = 352 K).
In any case, the spectral resolution and sensitivity of the present dataset is not
high enough to properly investigate this
intriguing possibility, calling for further 
interferometric observations. The analysis presented in this
paper only refers to the compact component.

\begin{figure*}
\begin{center}
\graphicspath{{figures/}}
\includegraphics[scale=0.69]{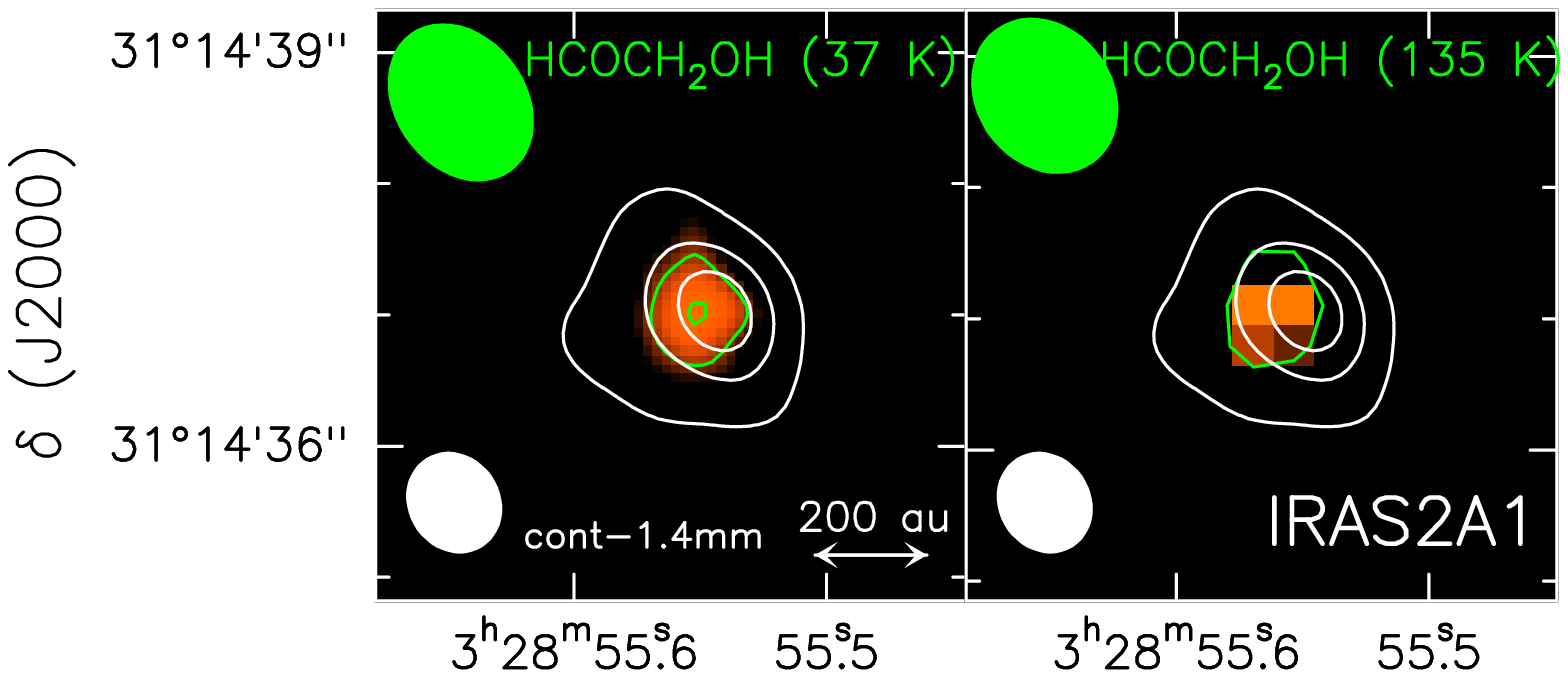} 
\includegraphics[scale=0.69]{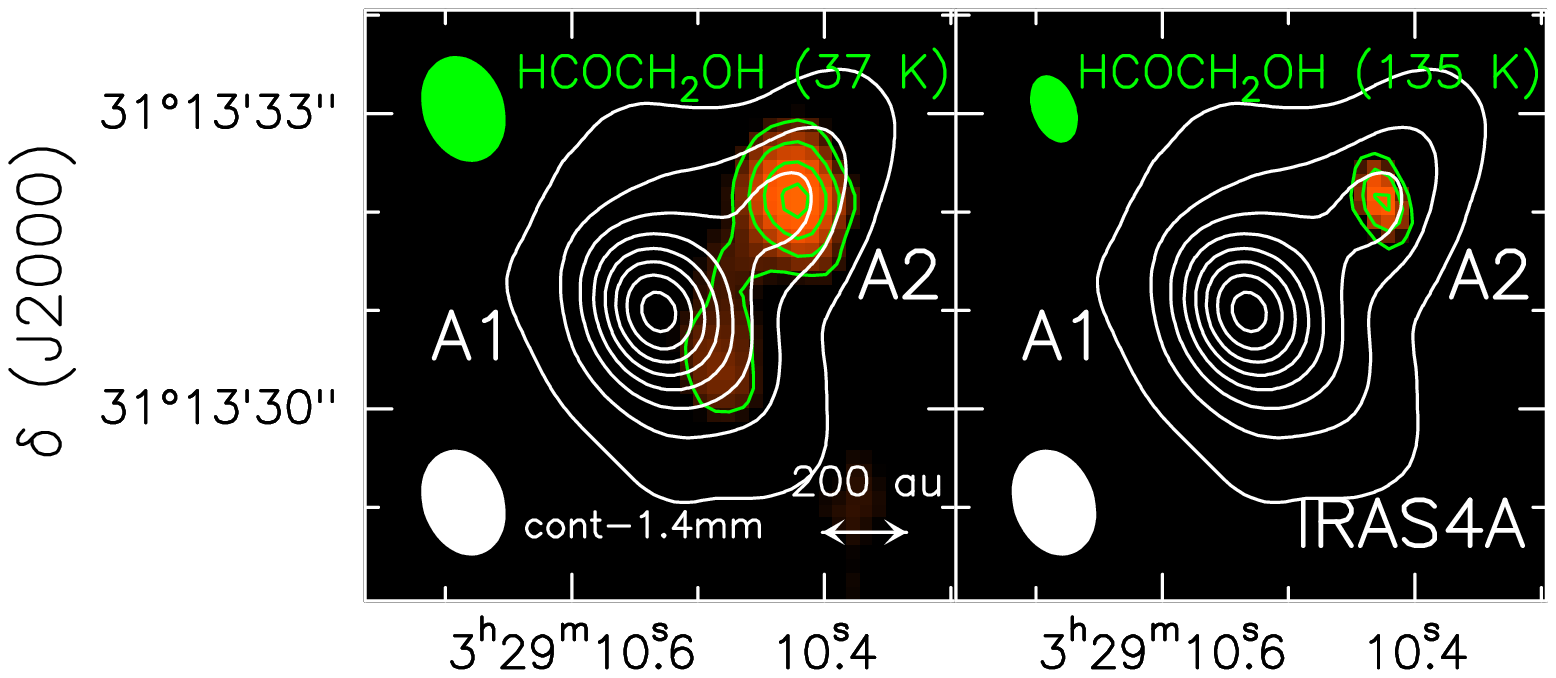}  
\includegraphics[scale=0.69]{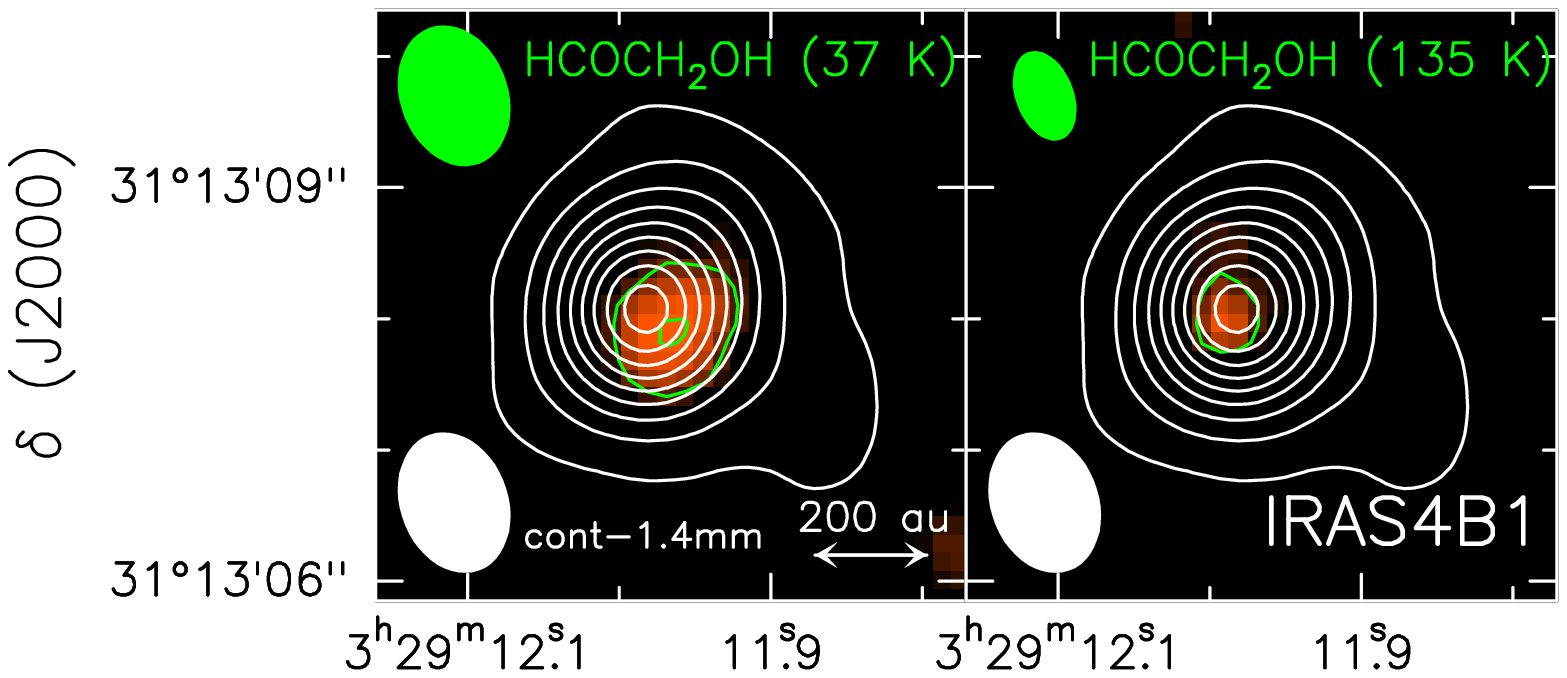} 
\includegraphics[scale=0.69]{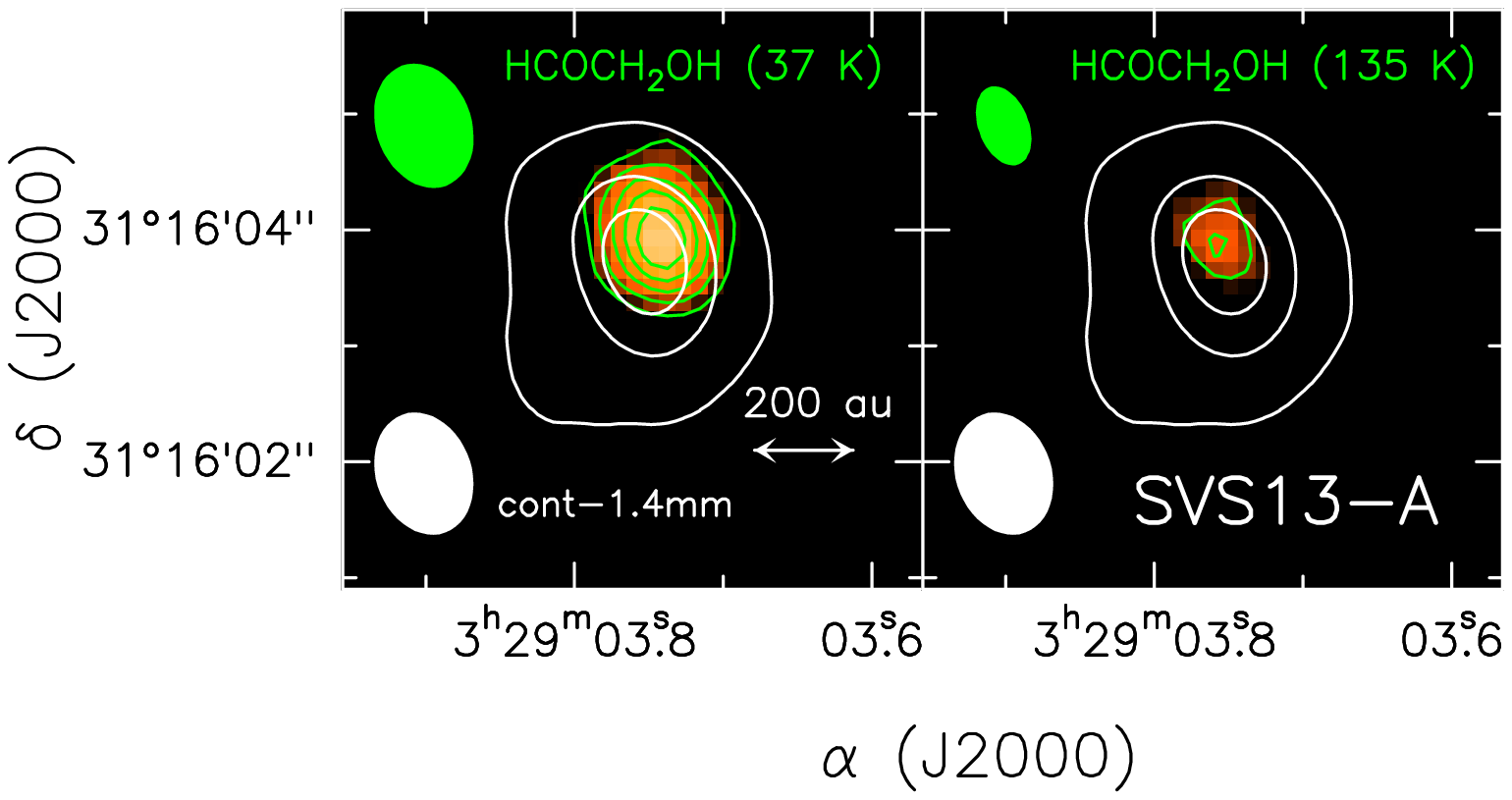}
\caption{Comparison between the 1.4mm continuum (white contours) 
and the glycolaldehyde spatial distributions (colour scale and green contours) 
towards (from top to bottom) NGC1333-IRAS2A1, NGC1333-IRAS4A, NGC1333-IRAS4B1, and
NGC1333 SVS13-A. For continuum, first contours and steps correspond to 5$\sigma$ 
(14 mJy beam$^{-1}$) and 10$\sigma$, respectively. The ellipses show
the PdBI synthesised beam (HPBW) for the continuum (white, see Table 1) and glycolaldehyde
images. The HPBWs for the glycolaldehyde images at 1.4mm are: 
0$\farcs$82$\times$0$\farcs$70, 32$\degr$ (NGC1333-IRAS2A1), 
1$\farcs$08$\times$0$\farcs$81, 19$\degr$ (NGC1333-IRAS4A),
1$\farcs$08$\times$0$\farcs$83, 20$\degr$ (NGC1333-IRAS4B1), and 
1$\farcs$06$\times$0$\farcs$80, 20$\degr$ (NGC1333 SVS13-A).
The HPBWs for the glycolaldehyde images at 1.3mm are: 
0$\farcs$82$\times$0$\farcs$70, 32$\degr$ (NGC1333-IRAS2A1), 
0$\farcs$72$\times$0$\farcs$44, 21$\degr$ (NGC1333-IRAS4A),
0$\farcs$72$\times$0$\farcs$44, 21$\degr$ (NGC1333-IRAS4B1), and 
0$\farcs$71$\times$0$\farcs$43, 21$\degr$ (NGC1333 SVS13-A).
{\it Left panels:} The glycolaldehyde distribution refers to
the sum of the 7$_{6,2}$-6$_{5,1}$ and 7$_{6,1}$-6$_{5,2}$ emission with (220.2 GHz, $E_{\rm u}$ = 37 K; 
see Tables A1-A4). First contours and steps correspond to 5$\sigma$ 
(90, 90, 75, and 55 mJy beam$^{-1}$ km s$^{-1}$ respectively for NGC1333-IRAS2A1, NGC1333-IRAS4A, 
NGC1333-IRAS4B1, and NGC1333 SVS13-A) and 3$\sigma$, respectively. 
{\it Right panels:} The glycolaldehyde distribution refers to
the 22$_{2,21}$-21$_{1,20}$ emission (232.3 GHz, $E_{\rm u}$ = 135 K; 
see Tables 2-5). First contours and steps correspond to 5$\sigma$ 
(100, 110, 50, and 60 mJy beam$^{-1}$ km s$^{-1}$ respectively for NGC1333-IRAS2A1, NGC1333--IRAS4A, 
NGC1333-IRAS4B1, and NGC1333 SVS13-A) and 3$\sigma$, respectively.} 
\end{center}
\end{figure*}

\subsection{HCOCH$_{2}$OH rotation diagrams}

\begin{figure*}
\begin{center}
\graphicspath{{figures/}}
\includegraphics[scale=0.4]{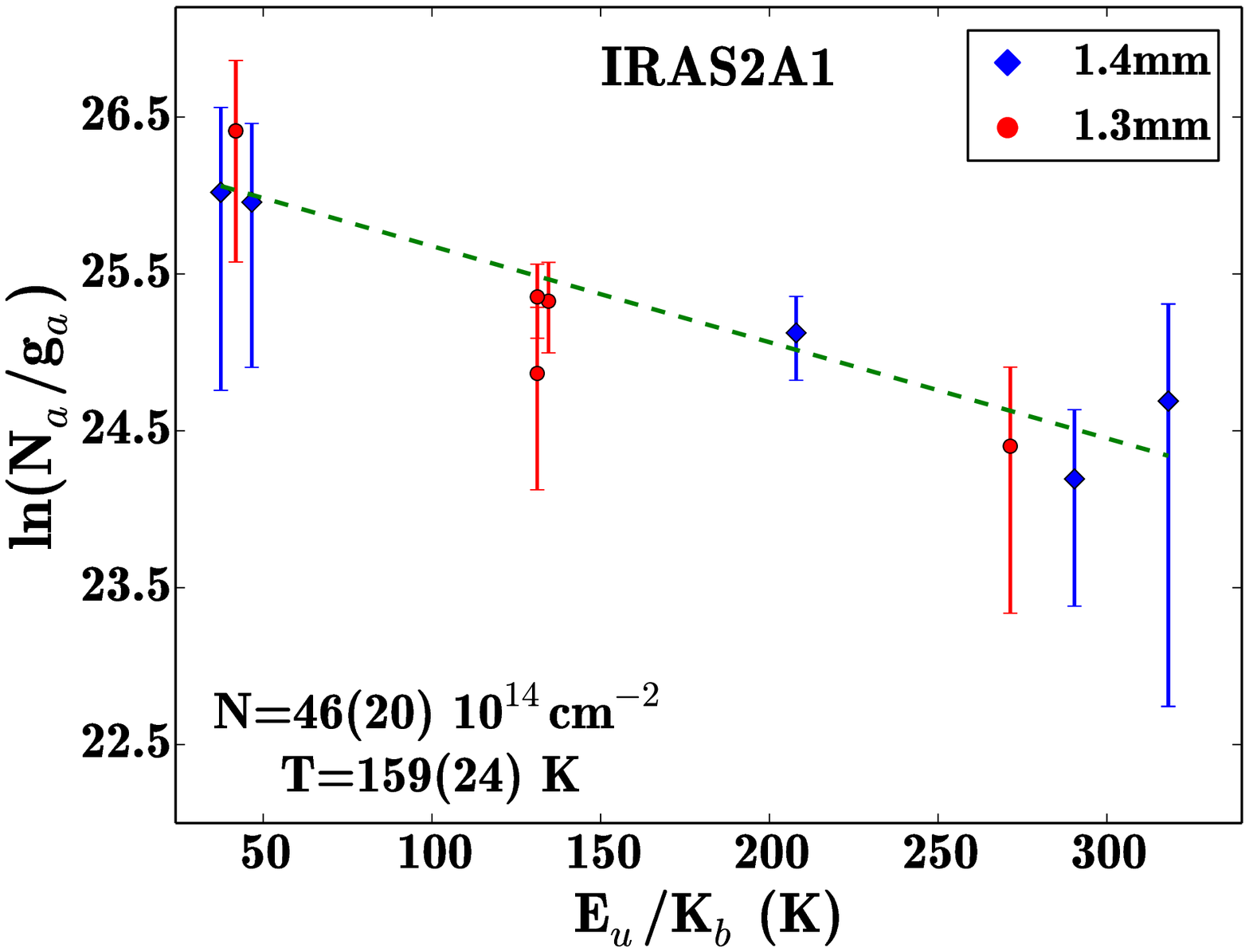} 
\includegraphics[scale=0.4]{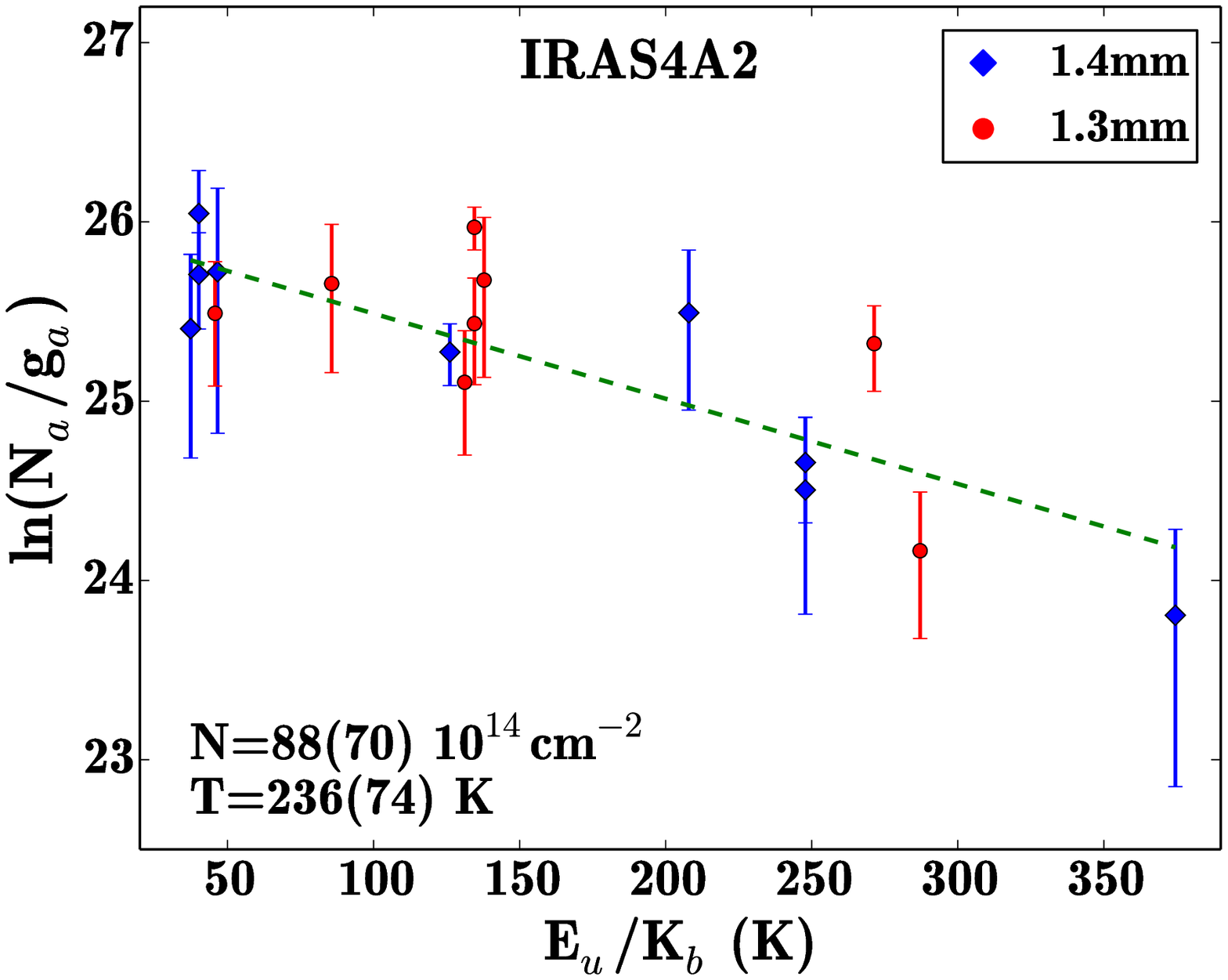} \\ 
\includegraphics[scale=0.4]{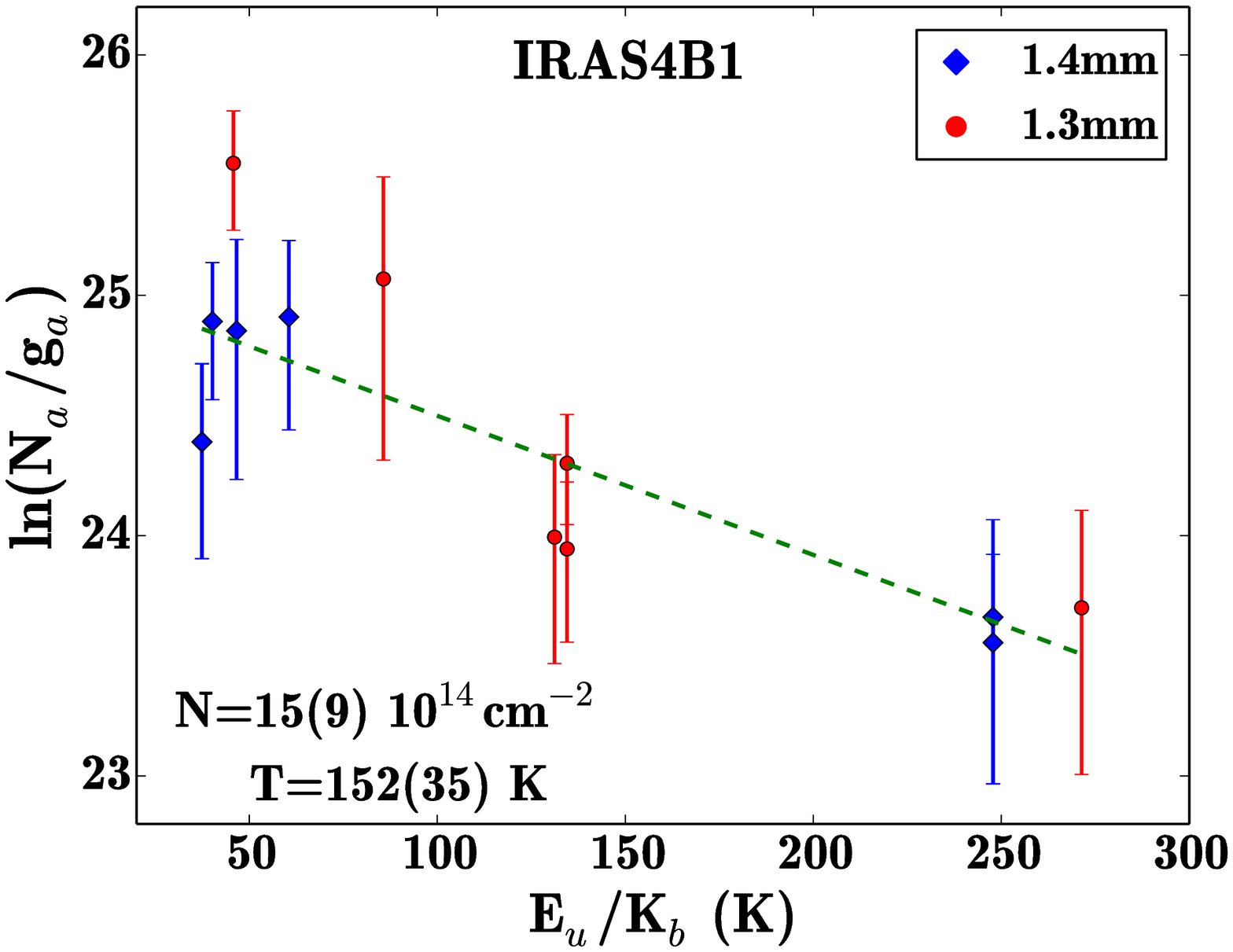} 
\includegraphics[scale=0.4]{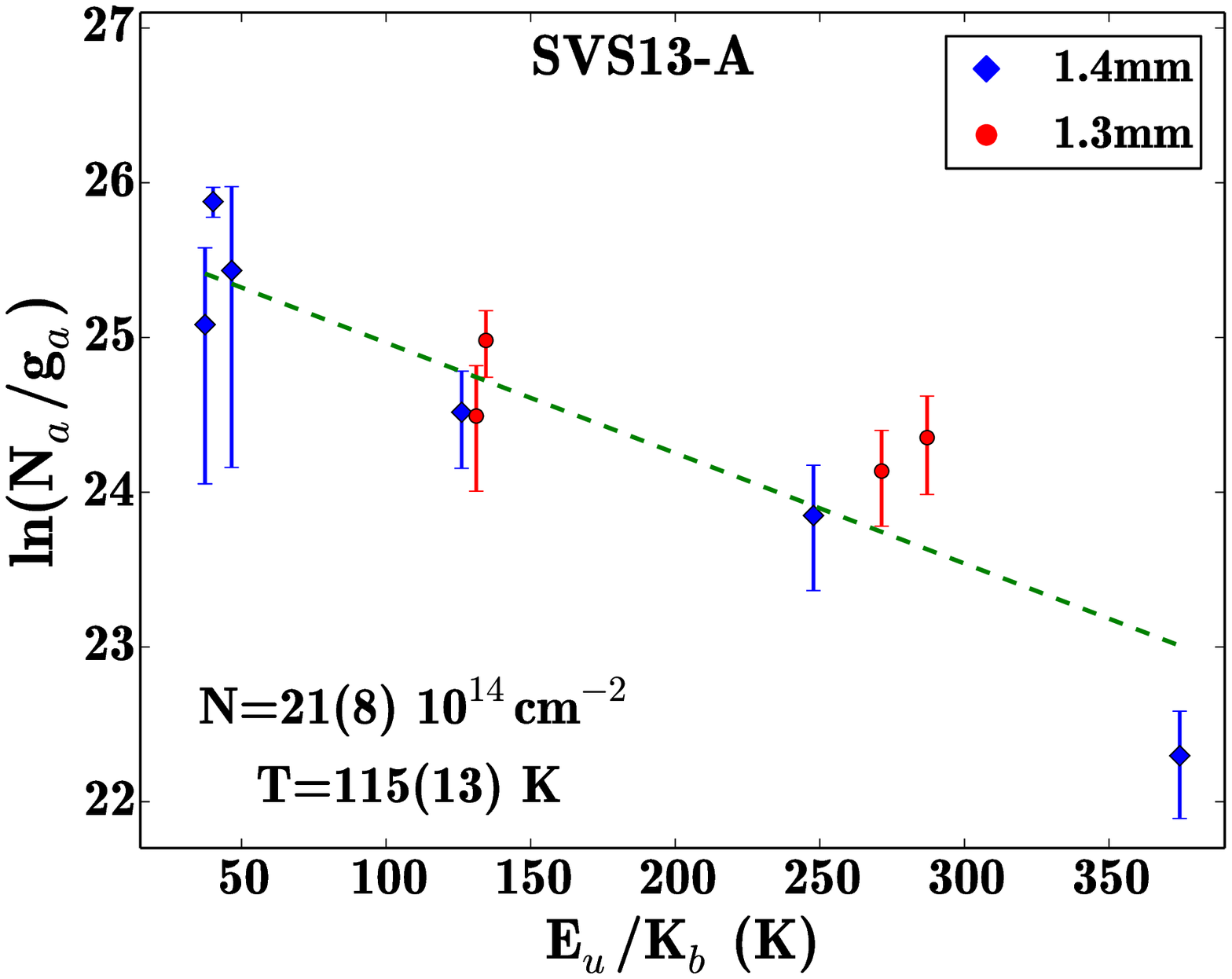}
\caption{Rotation diagram for the
glycolaldehyde transitions reported in Table 2
for NGC1333-IRAS2A ({\it Upper-left panel}), NGC1333-IRAS4A2
({\it Upper-right panel}), NGC1333-IRAS4B1 ({\it Lower-left panel}),
and NGC133-SVS13A ({\it Lower-right panel}). 
The line intensities have been corrected for beam dilution
using the sizes derived in Sect. 4.1. The parameters
$N_{\rm up}$, $g_{\rm up}$, and $E_{\rm up}$ are respectively
the column density, the degeneracy, and the energy of the upper level.
The error bars on ln($N_{\rm up}$/$g_{\rm up}$) are
given by the vertical bar of the symbols.
Red and blue points refer to the lines detected in the 
1.3 mm and 1.4mm spectral bands, respectively.
The dashed lines represent the best fits.}
\end{center}
\end{figure*}

Given that the glycolaldehyde collisional rates are not available in the literature, 
we derived excitation temperature and column densities by using a crude approach such as 
the classical rotational diagram, assuming a unique temperature, optically thin
emission, and Local Thermodynamic Equilibrium (LTE) conditions. We applied a correction due to the beam filling factor using the sizes reported in Sect. 4.1.
Under these assumptions, for a given transition the column density of the upper level, $N_{up}$, is related to the rotational temperature $T_{\rm rot}$, as follows:

\begin{equation}
N_{\rm up} = \frac{8\pi k\nu^2}{hc^3A_{\rm ul}ff} \int T dv 
\end{equation}

\begin{equation} 
ln \frac{N_{\rm up}}{g_{\rm up}} = ln N_{\rm tot} - ln Q(T_{\rm rot}) - \frac{E_{\rm up}}{k T_{\rm rot}} 
\end{equation}

where: $h$ and $k$ are the Planck and Boltzmann constants, respectively, $ff$ is the beam filling factor (derived using the sizes reported in Sect. 4.1), $g_{\rm up}$ is the degeneracy of the upper level, $N_{\rm tot}$ is the total column density, $Q$($T_{\rm rot}$) is the partition function, and $E_{\rm up}$ is the upper level energy. Figure 3 shows the rotation diagrams obtained using 
both the 1.3mm (red points) and 1.4mm (blue points) lines, while 
Table 2 reports the results, i.e. high temperatures, $T_{\rm rot}$ between 115 K and 236 K, and $N_{\rm tot}$
$\simeq$ 1--9 $\times$ 10$^{15}$ cm$^{-2}$. 
The present data
image, for the first time, a hot corino around SVS13-A and NGC1333-IRAS4B1.
These findings confirm that the glycolaldehyde gas is present in the 
region with a radius less than 100~au 
where the temperature is high enough to sublimate the dust mantles. Thus, either 
glycolaldehyde is directly released from mantle sublimation or formed through gas-phase reactions using simpler species directly formed on the mantles
and successively evaporated due to thermal heating.
Note that the sources where we report a detection of 
glycolaldehyde are those with the highest number of spectral lines of COMs 
(e.g. HCOOCH$_3$) detected above 6$\sigma$ in the 1.3mm and 1.4mm WideX spectra of the CALYPSO subsample of Perseus sources (Belloche et al., in preparation).
In the case of IRAS2A1 and IRAS4A2, the estimated rotational temperatures and column densities are well in agreement (after applying correctly the beam filling factor scaling) with what 
was previously reported by Taquet et al. (2015) and Coutens et al. (2015) using different
datasets on similar spatial scales, thus 
confirming the goodness of the present line identification. 
On the other hand, the derived $N_{\rm tot}$ are lower than
what J{\o}rgensen et al. (2012, 2016) towards
the IRAS16293-2422 binary 
(3--4 $\times$ 10$^{16}$ cm$^{-2}$) found, in Ophiuchus, on 
smaller ($\leq$ 60 au) spatial scales.
The sources where we report a detection of glycolaldehyde are those with the highest number of spectral lines of complex organic molecules detected above 6 sigma in the 1.3 mm and 1.4 mm WIDEX spectra of the CALYPSO subsample of Perseus sources (Belloche et al., in prep.).

The rotation diagram solutions have been used to create synthetic spectra 
(see the red lines in Fig. 1 and Figs. A.1 to A.9) to strengthen the 
detection using the GILDAS--Weeds package (Maret et al. 2011).
We assumed a background temperature of 2.73 K. In principle, for sources with a strong continuum such as NGC1333-IRAS4B1, 
the column density obtained using the rotational diagram method can be underestimated, 
converting the measurement in a lower limit.
The synthetic intensities are not systematically over- or
under-estimating the observed lines, confirming that 
the temperatures derived from the rotation diagrams
are good approximations. Note that all the detected lines are optically thin.
However, even though the rotational diagrams show reasonable agreement between
measurements and the single temperature model, it is worth to 
point out here
some shortcomings of our models in reproducing the observed spectra. 
While we do not find large discrepancies, the spectral resolution 
of our observations is relatively low and the spectra very rich of
emission lines (especially of the four sources where we do detect 
glycolaldehyde line). This implies that in some cases 
the lines may be contamined by emission from other molecules.
Figures A.1, A.3, A.5, and A.7 shows the lines selected for the analysis.
To be thorough, Figs. A.2, A.4, A.6, and A.8 show all the lines excluded from the present analysis because, although they are consistent with the observed spectrum, the blends with other species prevents a firm identification of the line. 
Note that, in a few cases, our simple LTE model predicts higher fluxes than observed; a special case is represented by the 20$_{2,18}$-19$_{3,17}$ line at 220463 MHz 
observed towards 
NGC1333-IRAS4B1, -IRAS2A1, IRAS4A2, and -SVS13A 
(see Fig. A.9), and conservatively excluded from the LTE analysis falling at the edge of the observed WideX band.
Beside this line, only in 2 cases (over a grand total of about 90 lines, counting also
those not used here because of blending) the discrepancy reaches a factor 3: 
the 22$_{1,21}-21_{2,20}$ line at $\sim$ 232286~MHz in IRAS2A1 (see Fig. A.2), and the 9$_{5,5}-8_{4,4}$ line at $\sim$
217626~MHz in SVS13-A (see Fig. A.8).
The difference is even less 
once considered both the errors associated with the LTE fit 
(see Table 2) and the errors on the observed line peak intensity.
Indeed, for lines close to the detection limit the 
systematic discrepancy could be understood in terms of the noise fluctuations and the uncertainty of the 
continuum subtraction in a region of the spectrum very rich of lines. We also 
note that deviations from the simplifying assumptions of single temperature 
LTE, non thermal excitation and absorption may be the cause of some of 
the discrepancies. Obtaining higher sensitivity and higher spectral 
and spatial resolution observations over a broader frequency range, 
together with a full spectrum modeling, will help deriving more accurate estimates 
of the gas parameters. Nonetheless, eventually the intrinsic linewidth, the source
chemical complexity and richness of the spectra and the source physical complexity 
will always limit the ability to accurately identify and measure individual lines. 
Overall, our simple LTE analysis and comparison with the spectra show that, while not 
perfect, it is possible to estimate a temperature and column density representative 
of the bulk of the molecular emission.

\begin{figure}
\begin{center}
\graphicspath{{figures/}}
\includegraphics[scale=0.4]{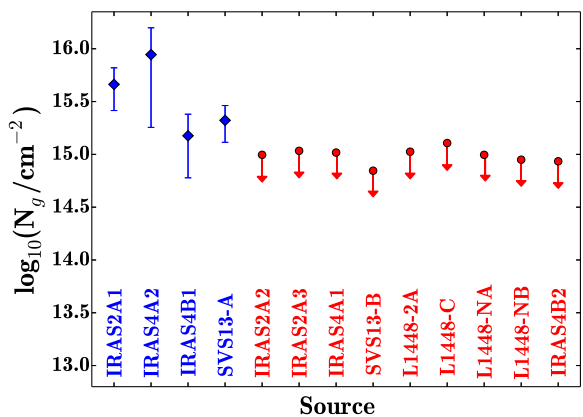}
\caption{Distribution of the glycolaldehyde column density ($N_{\rm tot}$) derived towards the selected sample (see Table 1). Red arrows are for upper limits.}
\end{center}
\end{figure}

\subsection{HCOCH$_{2}$OH occurrence around protostars}

Figure 4 compares the column densities measured towards
the four sources detected in glycolaldehyde with the upper limits on $N_{\rm tot}$
derived for the 9 objects in Perseus where glycolaldehyde has not been revealed. Assuming a typical temperature of 150 K, a size of 0$\farcs$4
(see Sect. 4.1), 
and considering the 3$\sigma$ values of the spectra where
no glycolaldehyde emission has been found, 
the typical upper limit is $\leq$ 10$^{15}$ cm$^{-2}$, i.e. one order of magnitude less than what found in IRAS2A1, IRAS4A2, IRAS4B1, and SVS13-A. 
These measurements support that these four protostars 
show a significantly higher column density of glycolaldehyde molecules
in the gas phase. However, this may not reflect directly a difference in
glycolaldehyde abundance.

To estimate the possible enhancement (or deficit) of gas phase glycolaldehyde abundance 
in the sources in our sample we need to compare with an estimate of the total gas column
density in each source.
As a proxy for the total column density towards each source we use the peak continuum flux at 1.4~mm.
The millimetre continuum flux in these sources is dominated by the thermal emission from dust
grains. The exact conversion of the millimeter flux into total gas column density depends
on three very uncertain factors: the dust properties (defining the dust opacity per gram of dust), the dust density and temperature structure, and the gas to dust ratio. Comparing the observed glycolaldehyde column densities with the continuum peak flux will allow to search for systematic abundance variations in our sample only under the assumption that the sources in our sample have uniform properties. While this is expected, since our sources are an homogeneous set of protostars in the same star forming region, in the future this analysis will have to be refined using more extensive observations and more accurate estimates of the total gas column density in each source.

Figure 5 shows 
the ratio between the glycolaldehyde column
density, $N_{\rm tot}$, and the continuum peak flux at 1.4~mm ($F_{\rm 1.4mm}$, from Maury et al. in preparation) against  the value of the 1.4~mm peak flux itself.
As mentioned above, if we assume that the molecular
hydrogen column density scales with the dust peak flux density, then the plot
in Fig. 5 shows that the gas phase abundance of glycolaldehyde relative to
molecular hydrogen has a spread of a factor of about 10 among the 
sources with glycolaldehyde detections. Even if the non-detection of sources with $F_{\rm 1.4mm}$  less than 100 mJy might suggest that a detection requires bright continuum emission, a bright continuum source is not necessarily associated with glycolaldehyde emission.
A typical example is given by the IRAS4-A1+A2 system,
where it is only the object showing weaker continuum emission that is
detected in glycolaldehyde (see Fig. 2). If we consider the sources with continuum peak flux above $\sim$100~mJy/beam, then the average ratios for sources with detected glycolaldehyde are a factor of $\sim$10 higher than the sources with upper limits. This suggests that the detected sources may have a significantly higher gas phase abundance of glycolaldehyde compared with the non-detected sources.

The glycolaldehyde detection indicates that the gas reached a temperature of
at least 100~K; at these high temperatures the gas is enriched chemically
by the release of molecules directly from the dust mantles and/or by
the overcome of the activation barrier of gas phase reactions
using what was released from the icy dust mantles. 
It is reasonable to assume that the temperature increases with the source 
luminosity, which depends on the protostellar mass and on the source accretion rate, which in turn 
is related to the evolutionary stage.
However, only three sources where glycolaldehyde has been detected can be associated with a
measurement of the internal luminosity hampering a reliable check for a trend (see Table 1).
In addition, our 1.4mm fluxes do not allow us to derive a direct measurement of the protostellar mass. 
Trying to shed light on this aspect, we note that in the case of the SVS13-A+B system, 
glycolaldehyde is only detected towards the more evolved SVS13-A source, 
with $L_{\rm int}$ = 24.5 $L_{\rm \sun}$,
whereas the younger object SVS13-B, with $L_{\rm int}$ = 1--2 $L_{\rm \sun}$.
does not show evidence of an enriched chemistry.
Therefore, at least in this case, the lack of a detected hot-corino in SVS13-B suggests that the protostar 
had not enough time to warm a portion of gas large/bright enough to be detectable. 
Interestingly, Santangelo et al. (2015), using the SiO and CO jet properties imaged in IRAS4-A1+A2 system, proposed that the chemical richness observed towards IRAS4-A2 (and not towards IRAS4-A1),
is due to a later evolutionary stage, inferred from the jet properties driven by the two objects.
Nevertheless, it is clear that a more systematic comparative study of sources in different evolutionary phases is required to reach a firm conclusion:
based only on a few sources (e.g. SVS13-A+B), it is 
very difficult to tell whether the primary factor leading to the detection of 
glycolaldehyde emission is evolution (e.g. low value of $L_{\rm submm}$/$L_{\rm int}$) or accretion 
luminosity (high $L_{\rm int}$) or environment (e.g. NGC1333 vs. L1448).  
These findings call for further analysis of the chemical content of the protostellar environments of the whole CALYPSO sample, using all the emission due to all the COMs and not only to glycolaldehyde, which will be presented in a forthcoming paper (Belloche et al., in preparation).

\begin{figure}
\begin{center}
\graphicspath{{figures/}}
\includegraphics[scale=0.4]{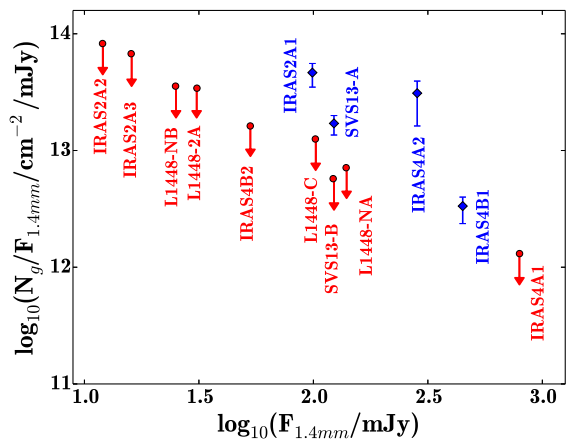} 
\caption{Ratio between the glycolaldehyde column density ($N_{\rm tot}$) and the peak flux of the 1.4mm continuum
emission ($F_{\rm 1.4mm}$; Maury et al., in preparation) versus the 1.4mm peak flux. Red arrows are for sources where glycolaldehyde has not been detected.}
\end{center}
\end{figure}

\section{Conclusions}

We report a survey of glycolaldehyde emission towards a sample
of Class 0 protostars located in the L1448 and NGC1333 star forming regions, in Perseus.
The analysis has been performed in the framework of the IRAM PdBI CALYPSO survey.
The main findings can be summarised as follows:

\begin{enumerate}

\item

We found glycolaldehyde towards 3 Class 0 objects (NGC1333-IRAS2A1, NGC1333-IRAS4A2,
NGC1333-IRA4B1) and 1 Class I source (SVS13-A). 
Once excluded the continuum peaks whose nature is still debated, this results 
in a detection rate of $\sim$ 36\%. The NGC1333 star forming region is
particularly rich in glycolaldehyde, with a definitely higher rate of occurrence: up to 60\%.

\item
The glycolaldehyde emission is compact and coincident with the continuum peak
of each detected source. The brightest lines indicate typical
sizes of 0$\farcs$4 showing that the glycolaldehyde emission 
is tracing the inner 100~au around the protostar.
We detected a large (up to 18) number of lines, covering a large
range of upper-level energy, with  $E_{\rm u}$ from 37 K up to 375 K,
and indicating high temperatures, $T_{\rm rot}$ between 100 K and 200 K.
We confirm and extend earlier results towards IRAS2A1 and IRAS4A2, and 
reveal for the first time the presence of hot corinos around
NGC1333-IRAS4B1 and SVS13-A.
The total column densities of glycolaldehyde in the detected sources are $\simeq$ 1--9 $\times$ 10$^{15}$ cm$^{-2}$. 

\item
We suggest that the four hot corinos imaged in glycolaldehyde in NGC1333 are associated with
a significantly higher gas phase abundance of glycolaldehyde with respect to the rest of
the sample. The SVS13-A+B and IRAS4-A1+A2 systems suggest that the detection of glycolaldehyde emission
is due to the source evolution: the chemical richness highlights those
protostars evolved enough to warm a bright region of $\sim$100~au at 
temperatures larger than 100~K. However, the present sample does not allow us to
reach a firm conclusion whether the glycolaldehyde detection is always a question of evolution or 
accretion luminosity and/or if it depends on the environment
hosting the protostars.

\end{enumerate}

Our initial results can be confirmed and extended with more sensitive 
observations of a larger sample of objects. In particular, more accurate
determinations of the possible variations of glycolaldehyde abundance in gas phase
will be a powerful probe of the chemical evolution of protostars and may confirm the present results.

\begin{acknowledgements}
We are very grateful to all the IRAM staff, whose dedication allowed us to carry out the CALYPSO project.
This work was partly supported by the PRIN INAF 2012 -- JEDI and by the Italian Ministero dell'Istruzione, Universit\`a e Ricerca through the grant Progetti Premiali 2012 -- iALMA (CUP C52I13000140001).
AM is supported by the MagneticYSOs, grant agreement n.679937.
CC, LT, and AM have been partially supported by the
Gothenburg Centre for Advanced Studies in Science and Technology as part of the GoCAS program {\it Origins of Habitable Planets}.
The research leading to these results has received funding from the European Community’s Seventh Framework Programme (FP7/2007-2013/) under grant agreements No 229517 (ESO COFUND) and No 291294 (ERC ORISTARS), and from the French Agence Nationale de la Recherche (ANR), under reference ANR-12-JS05-0005.
\end{acknowledgements}

\clearpage

\appendix

\section{Observed HCOCH$_2$OH emission lines}

Tables A.1 to A.4 report, for the 4 sources where glycolaldehyde
emission has been detected (IRAS2A, SVS13-A, IRAS4A2, and IRAS4B), 
the results of the Gaussian fit for all the detected lines. 
In particular, we first list the spectral parameters (frequency, $\nu$, upper level energy, $E_{\rm u}$,
and the $S\mu^{2}$ product, all taken from the Jet
Propulsion Laboratory database; Pickett et al. 1998), followed by the
line parameters (in $T_{\rm MB}$ scale), namely: the integrated emission ($\int$ $T dv$), 
both peak line velocities ($V_{peak}$) and intensities ($T_{\rm peak}$), linewidth ($FWHM$),
and r.m.s. noise. 
Figures A.1, A.3, A.5, and A.7
show the lines detected and used for the analysis, Figs. A.2, A.4, A.6, and A.8
show the lines excluded from the analysis given the severe blending
with other emission lines, while Fig. A.9 is for a line which has
not been used because it is located at the edge of the WideX bandwidth.

\begin{figure*}
\begin{center}
\graphicspath{{figures/}}
\includegraphics[scale=0.52]{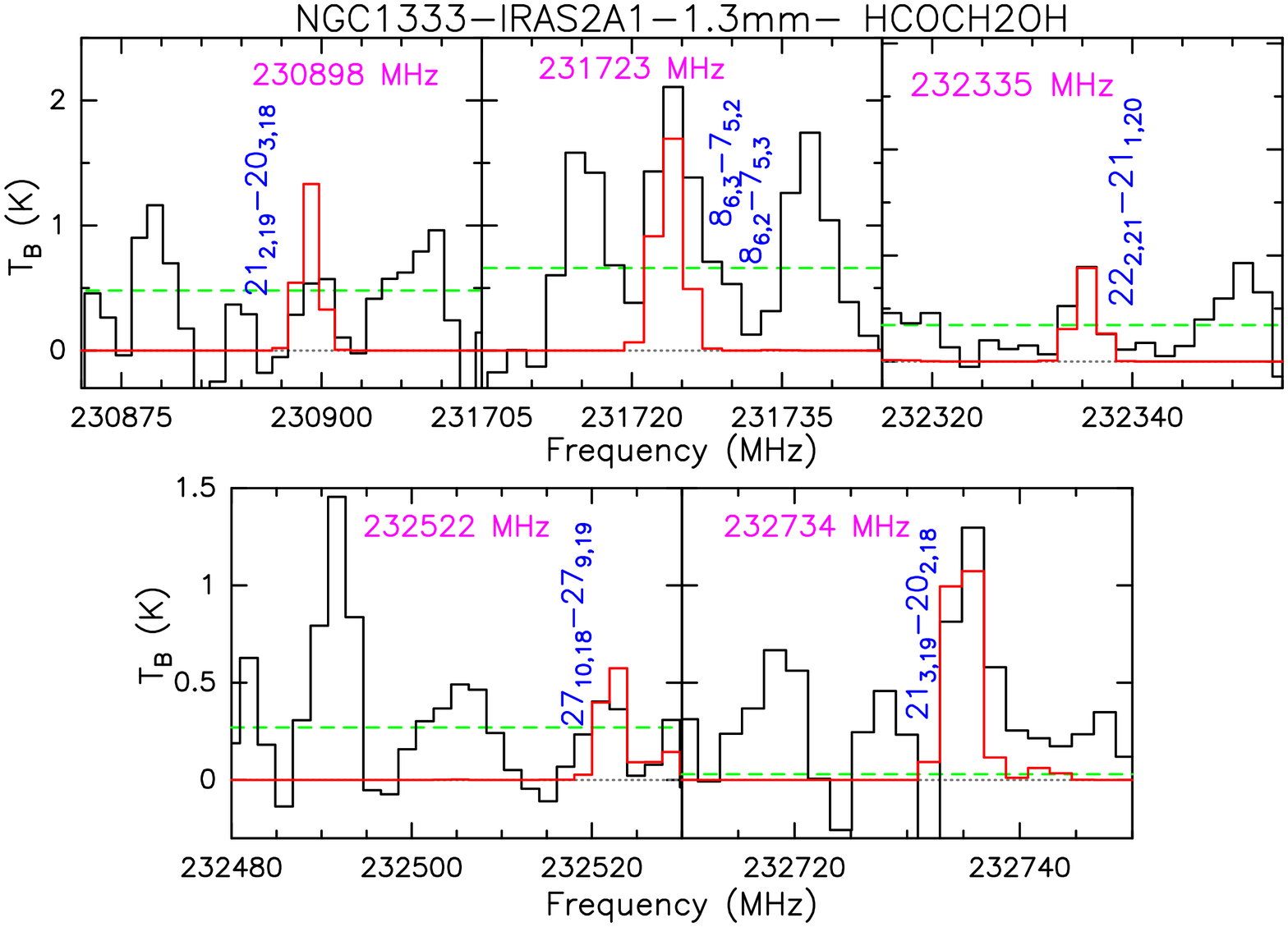} 

\includegraphics[scale=0.52]{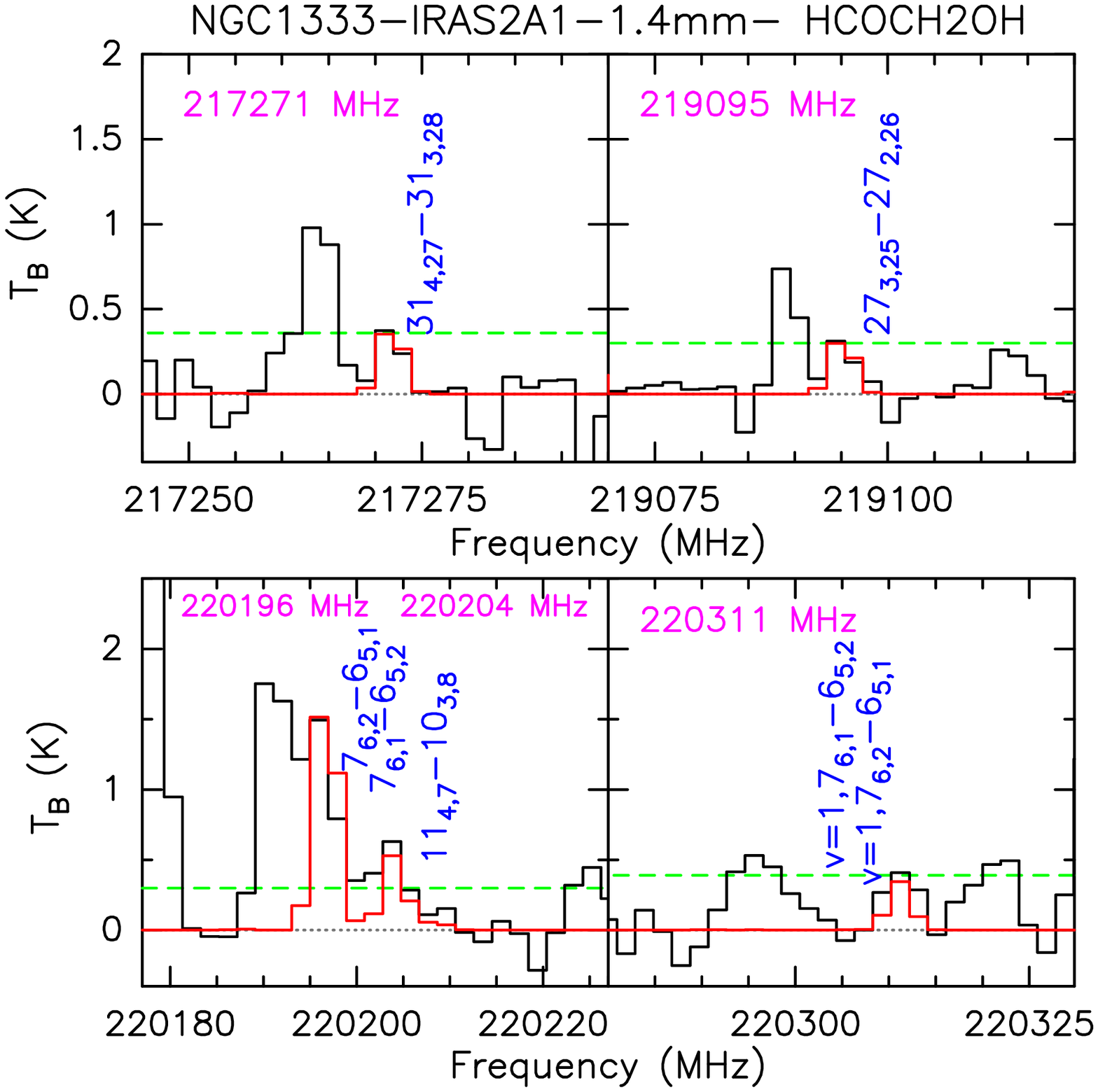}  
\caption{Glycolaldehyde emission lines (in $T_{\rm B}$ scale)
detected in the 1.3mm and 1.4mm spectral windows towards 
NGC1333-IRAS2A1 and used to perform the LTE analysis. 
The horizontal green dotted lines show the 3$\sigma$ level.
In blue we mark the glycolaldehyde lines extracted from Table A.1.
The red line shows the synthetic spectrum obtained with the GILDAS--Weeds package 
(Maret et al. 2011) and assuming the rotation diagram solutions (see Fig. 3).}
\end{center}
\end{figure*}

\begin{figure*}
\begin{center}
\graphicspath{{figures/}}
\includegraphics[scale=0.52]{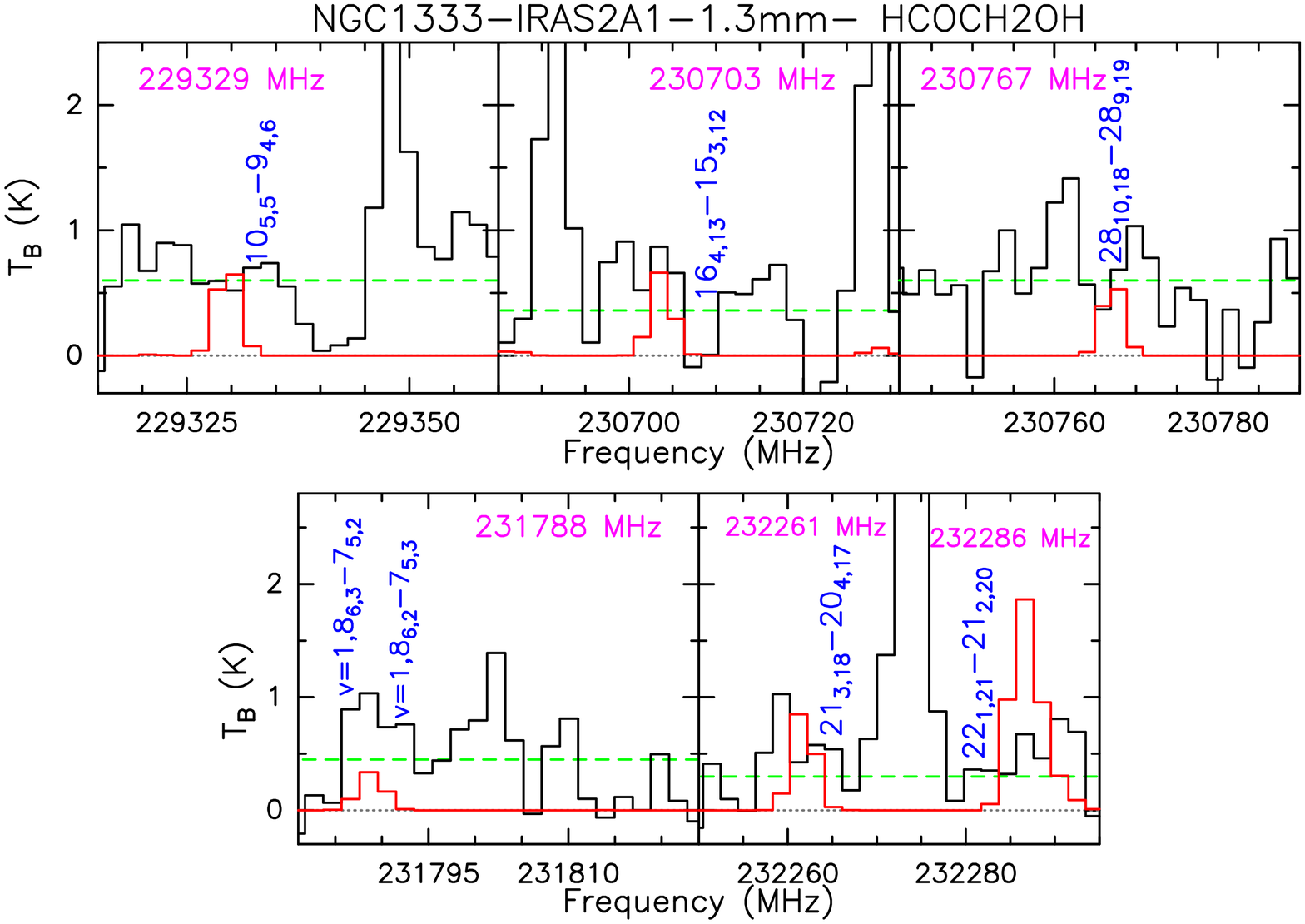} 
\includegraphics[scale=0.52]{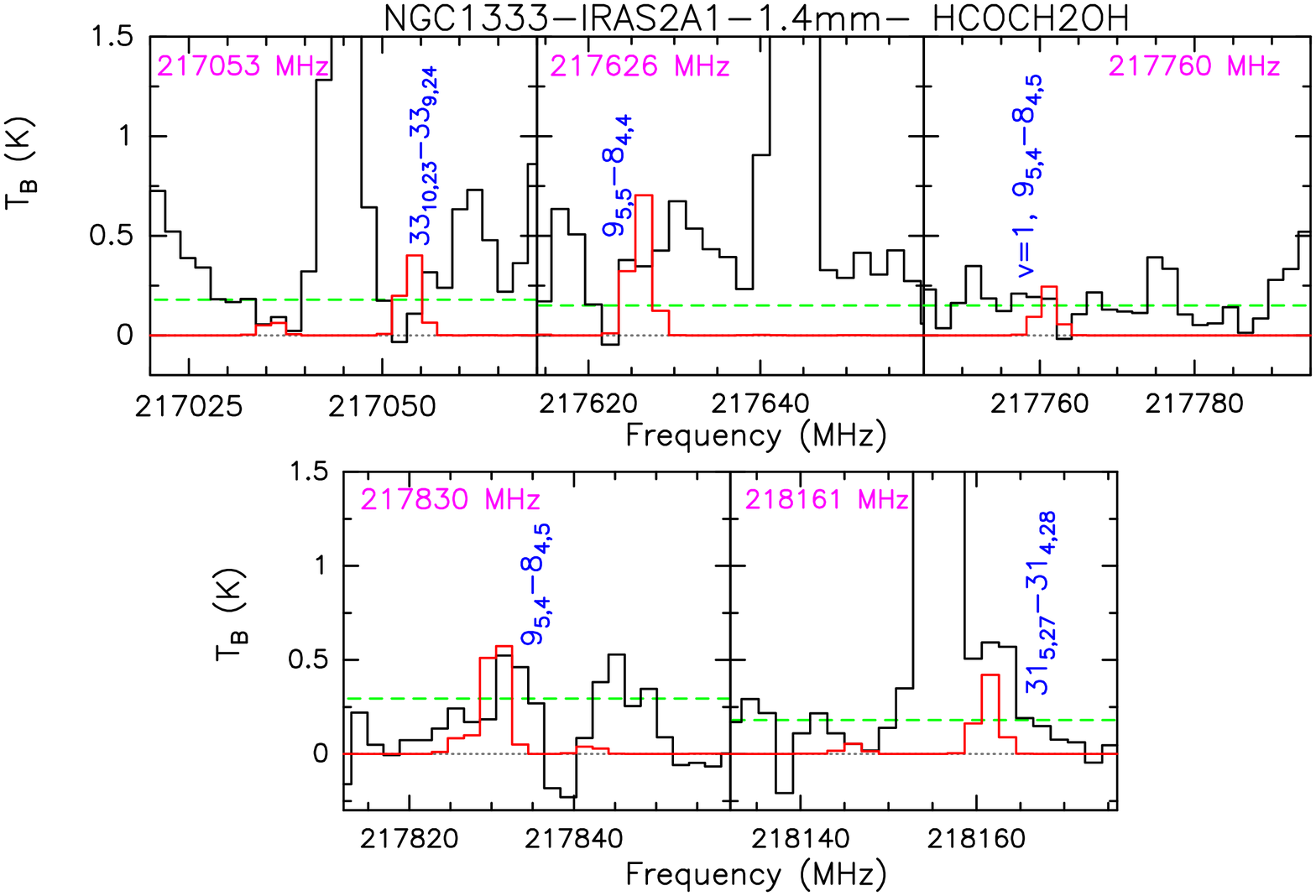}
\caption{Glycolaldehyde emission lines 
(in $T_{\rm B}$ scale) observed in the 1.3mm and 1.4mm spectral windows towards 
NGC1333-IRAS2A1 and excluded from the LTE analysis because of 
severe blending with other emission lines. 
The horizontal green dotted lines show the 3$\sigma$ level.
In blue we mark the glycolaldehyde transitions.
The red line shows the synthetic spectrum obtained with the GILDAS--Weeds package 
(Maret et al. 2011) and assuming the rotation diagram solutions (see Fig. 3).}
\end{center}
\end{figure*}

\begin{figure*}
\addtocounter{figure}{-1}
\begin{center}
\graphicspath{{figures/}}
\includegraphics[scale=0.52]{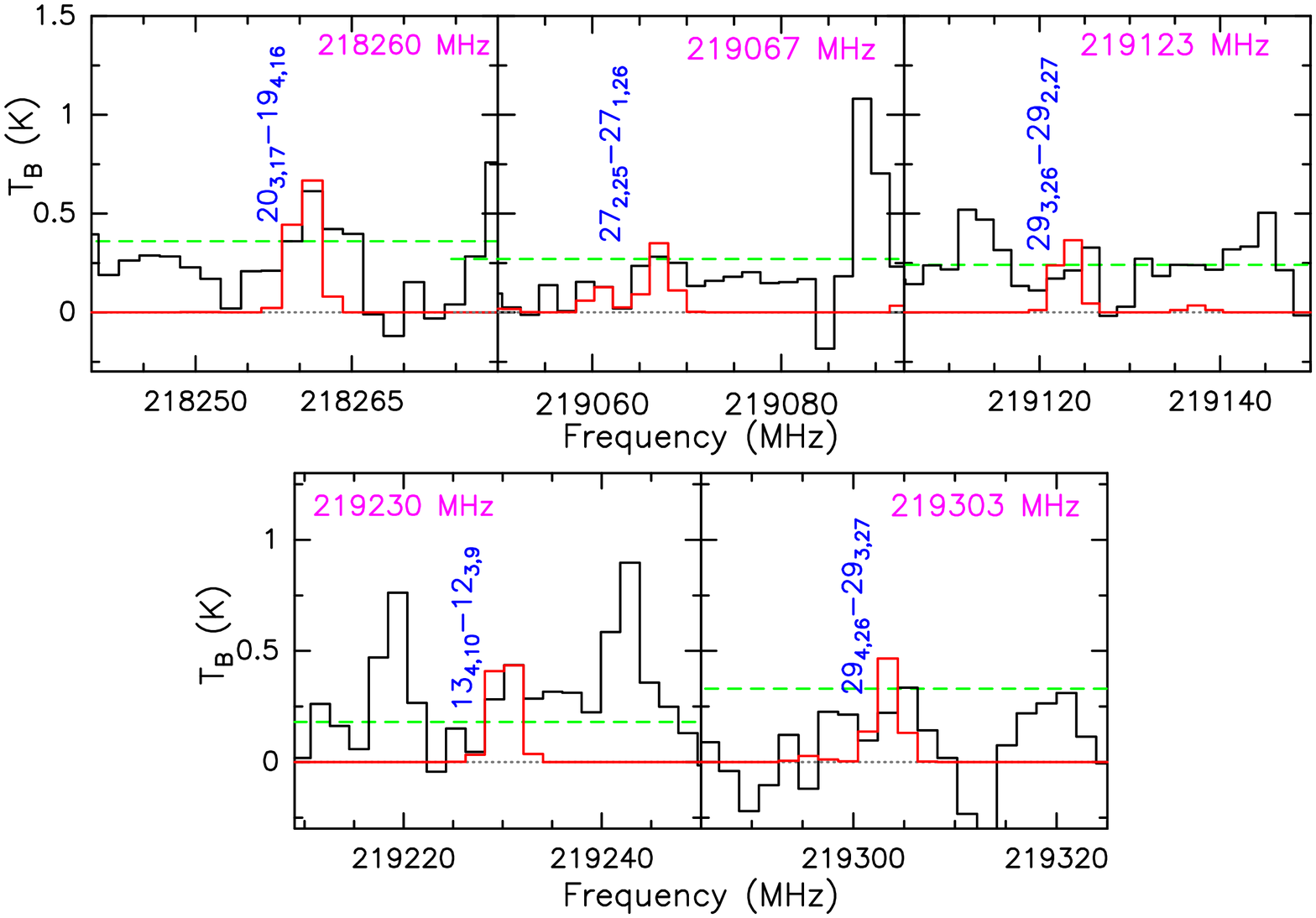}
\caption{{\it Continued}.}
\end{center}
\end{figure*}

\begin{figure*}
\begin{center}
\graphicspath{{figures/}}
\includegraphics[scale=0.52]{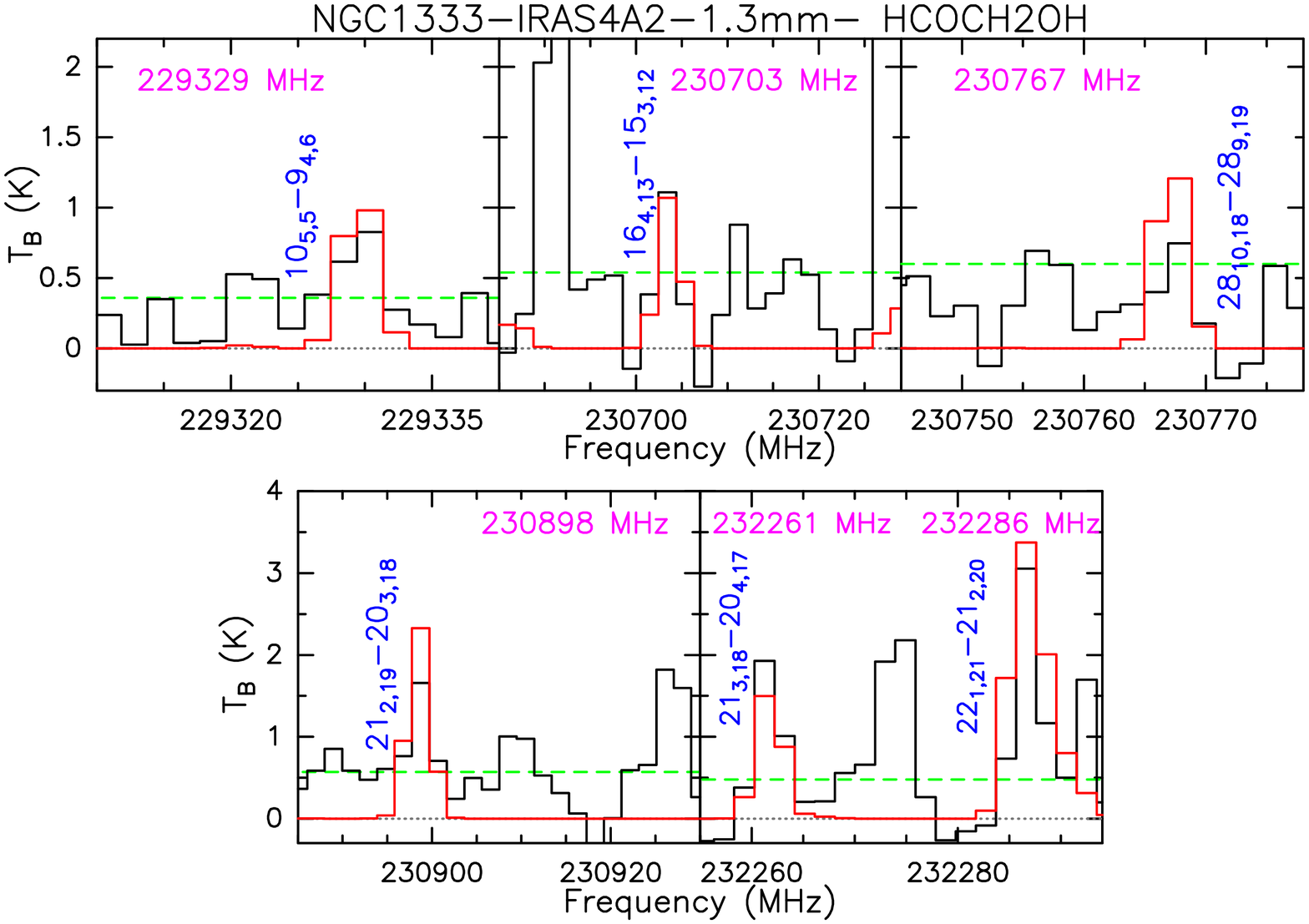} 
\includegraphics[scale=0.52]{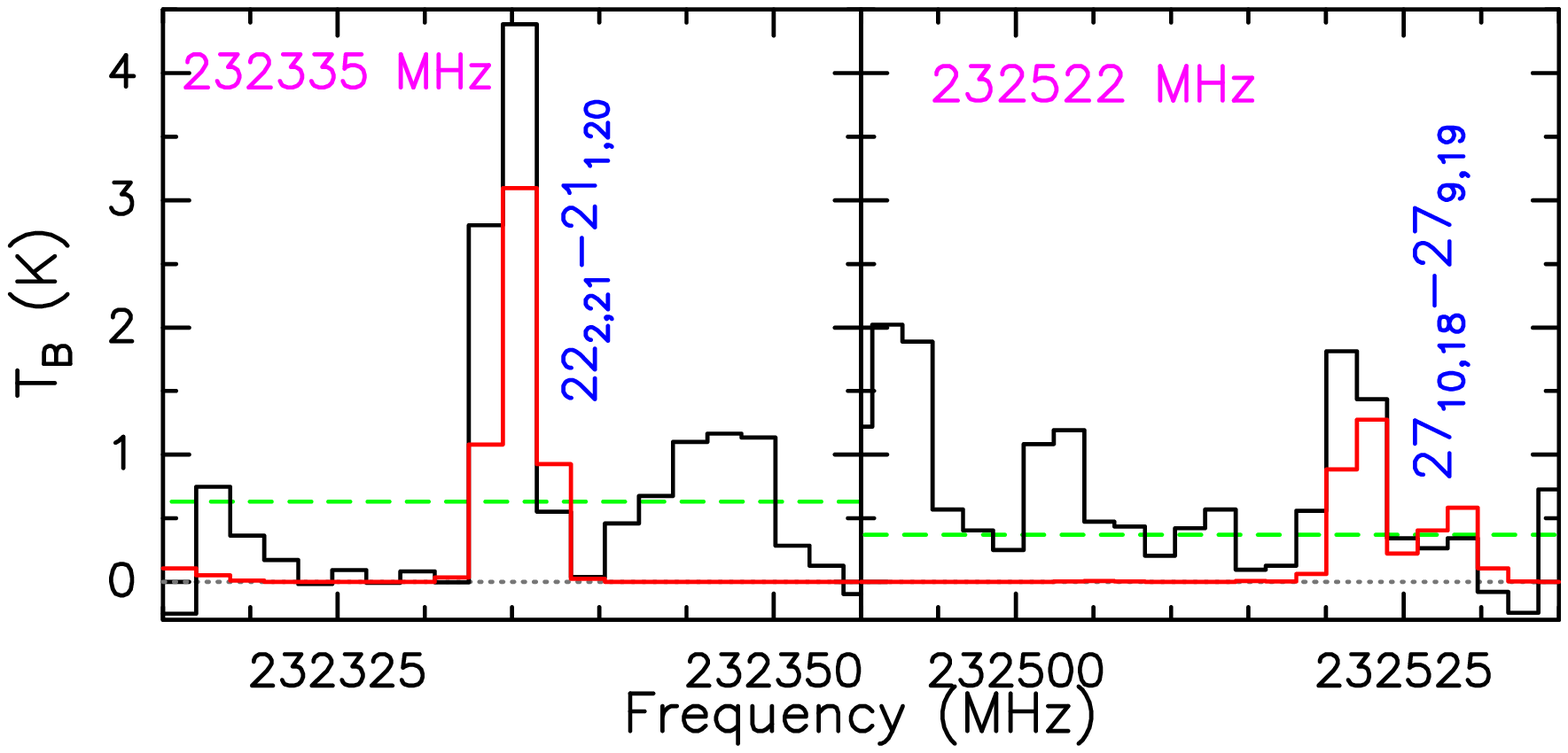} 
\caption{Glycolaldehyde emission lines (in $T_{\rm B}$ scale)
detected in the 1.3mm and 1.4mm spectral windows towards 
NGC1333-IRAS4A2 and used to perform the LTE analysis. 
The horizontal green dotted lines show the 3$\sigma$ level.
In blue we mark the glycolaldehyde lines extracted from Table A.2.
The red line shows the synthetic spectrum obtained with the GILDAS--Weeds package 
(Maret et al. 2011) and assuming the rotation diagram solutions (see Fig. 3).}
\end{center}
\end{figure*}

\begin{figure*}
\addtocounter{figure}{-1}
\begin{center}
\graphicspath{{figures/}}
\includegraphics[scale=0.5]{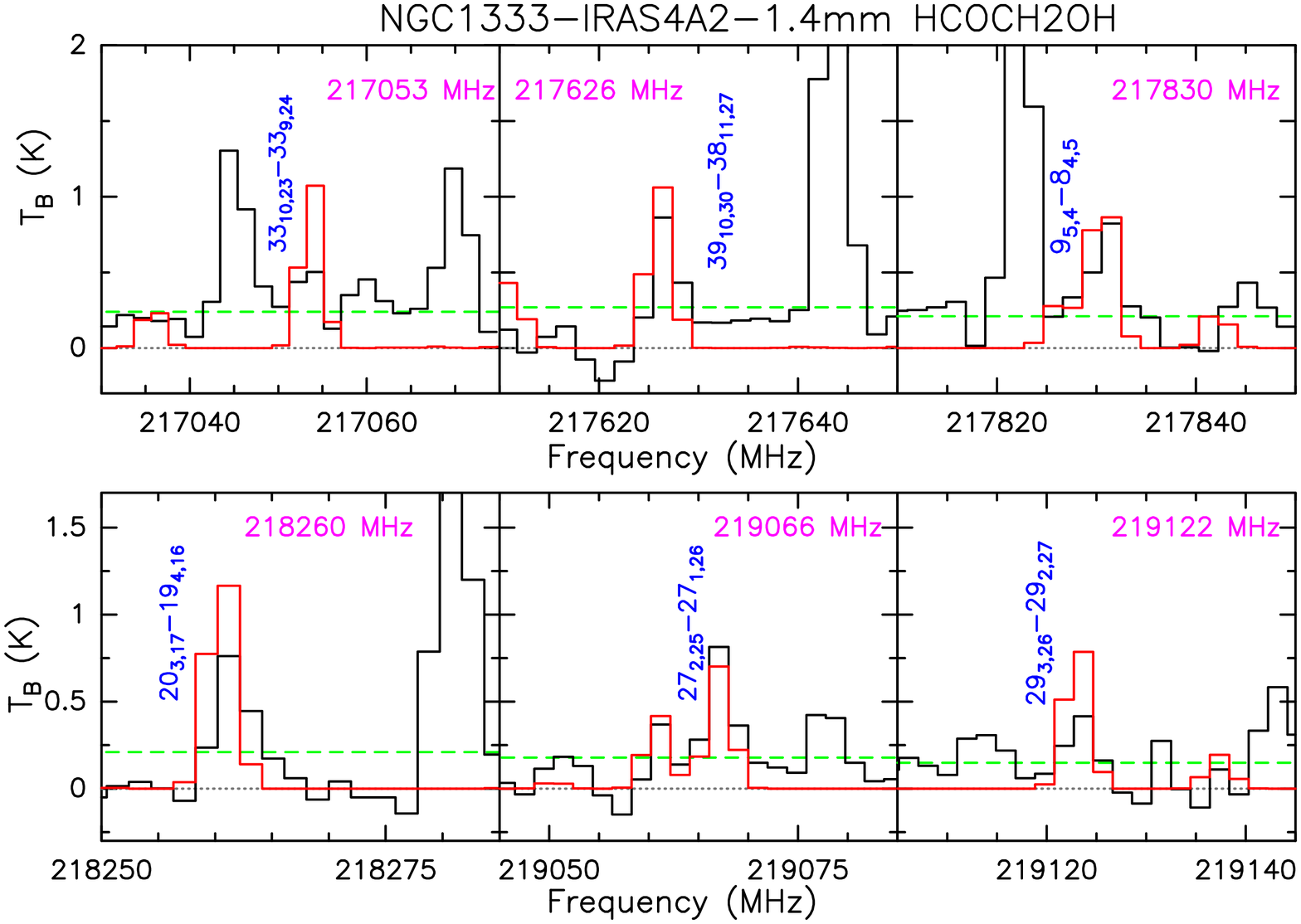}  
\includegraphics[scale=0.5]{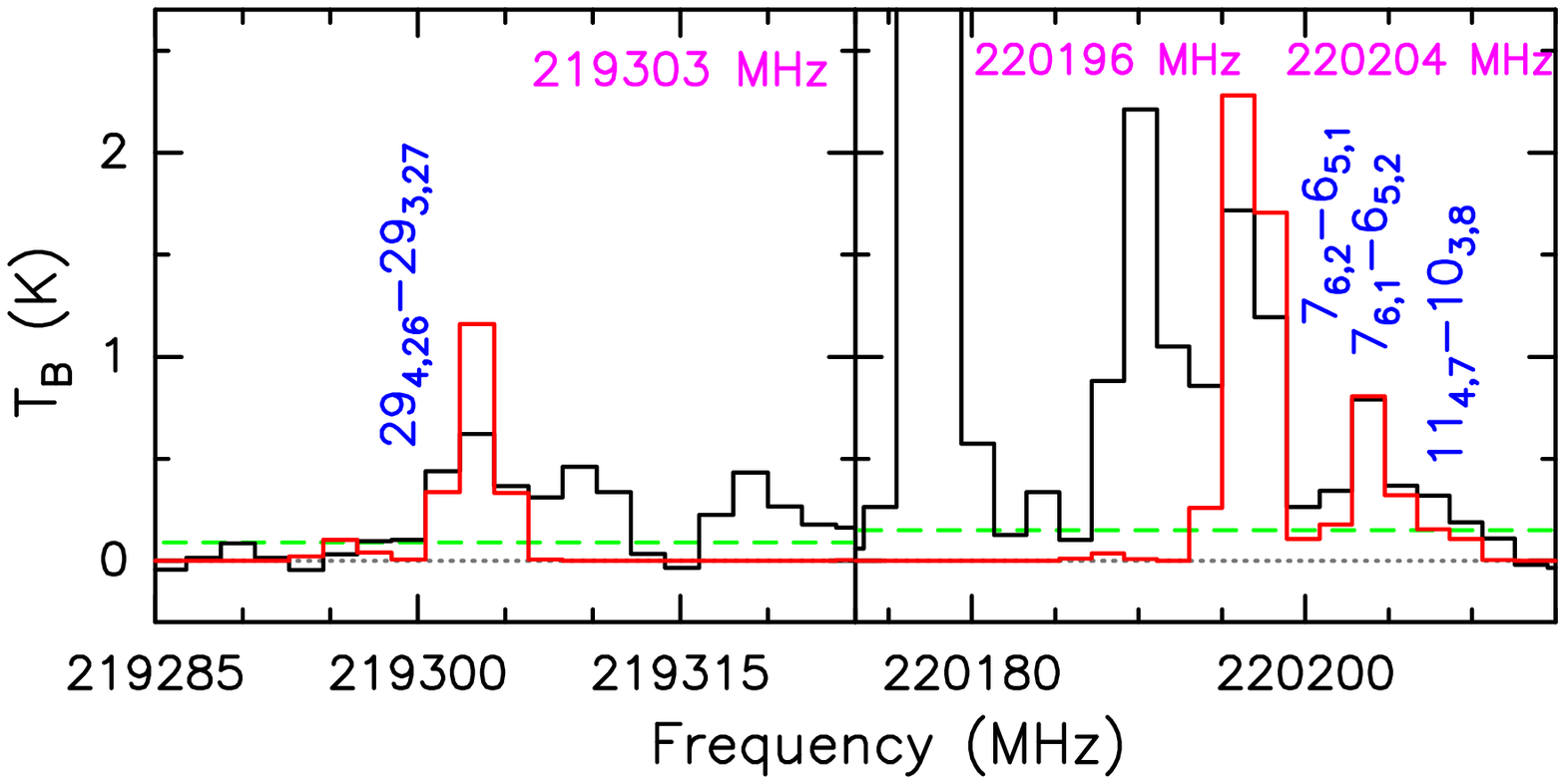}  
\caption{\it Continued.}
\end{center}
\end{figure*}

\begin{figure*}
\begin{center}
\graphicspath{{figures/}}
\includegraphics[scale=0.48]{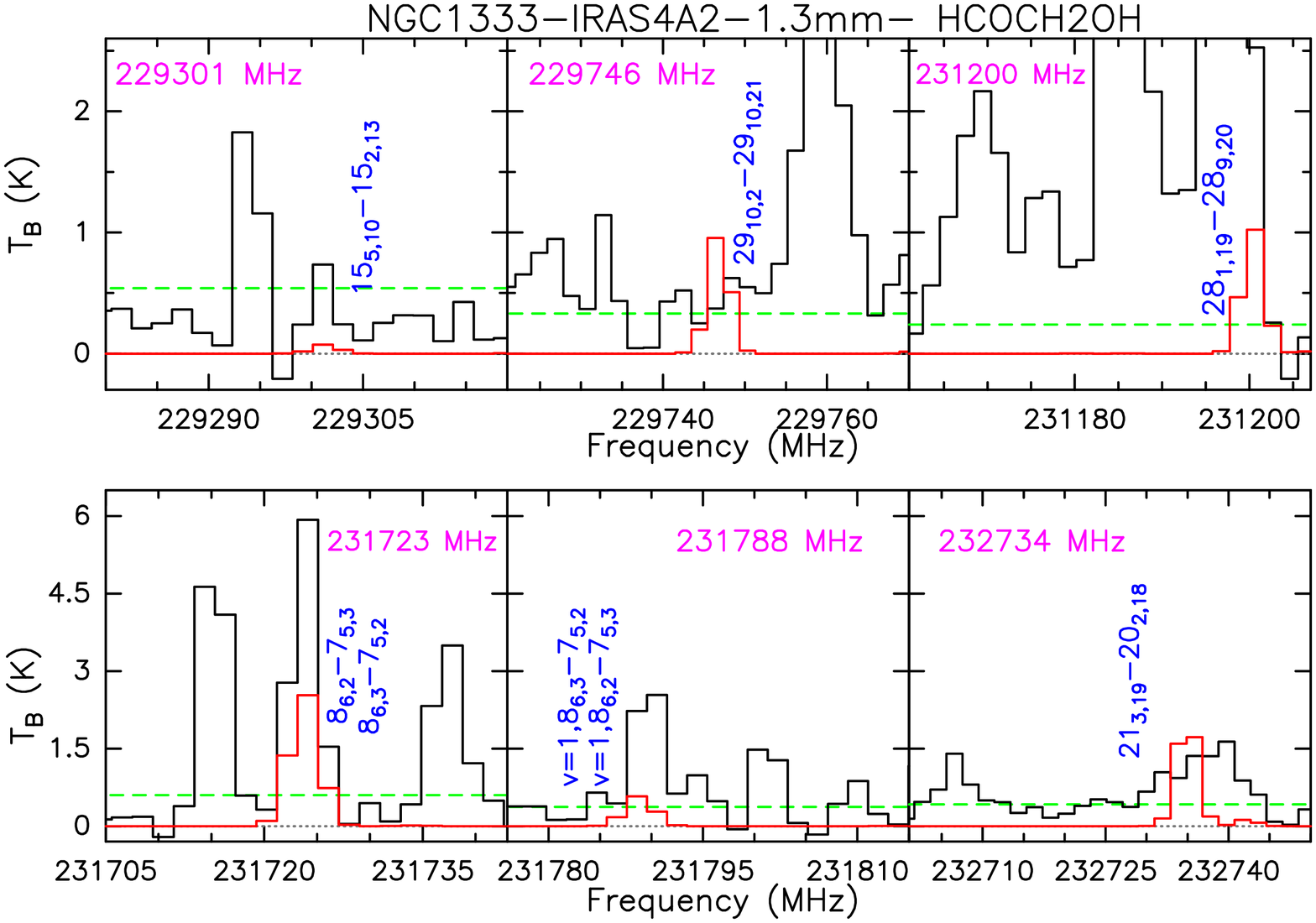}
\includegraphics[scale=0.48]{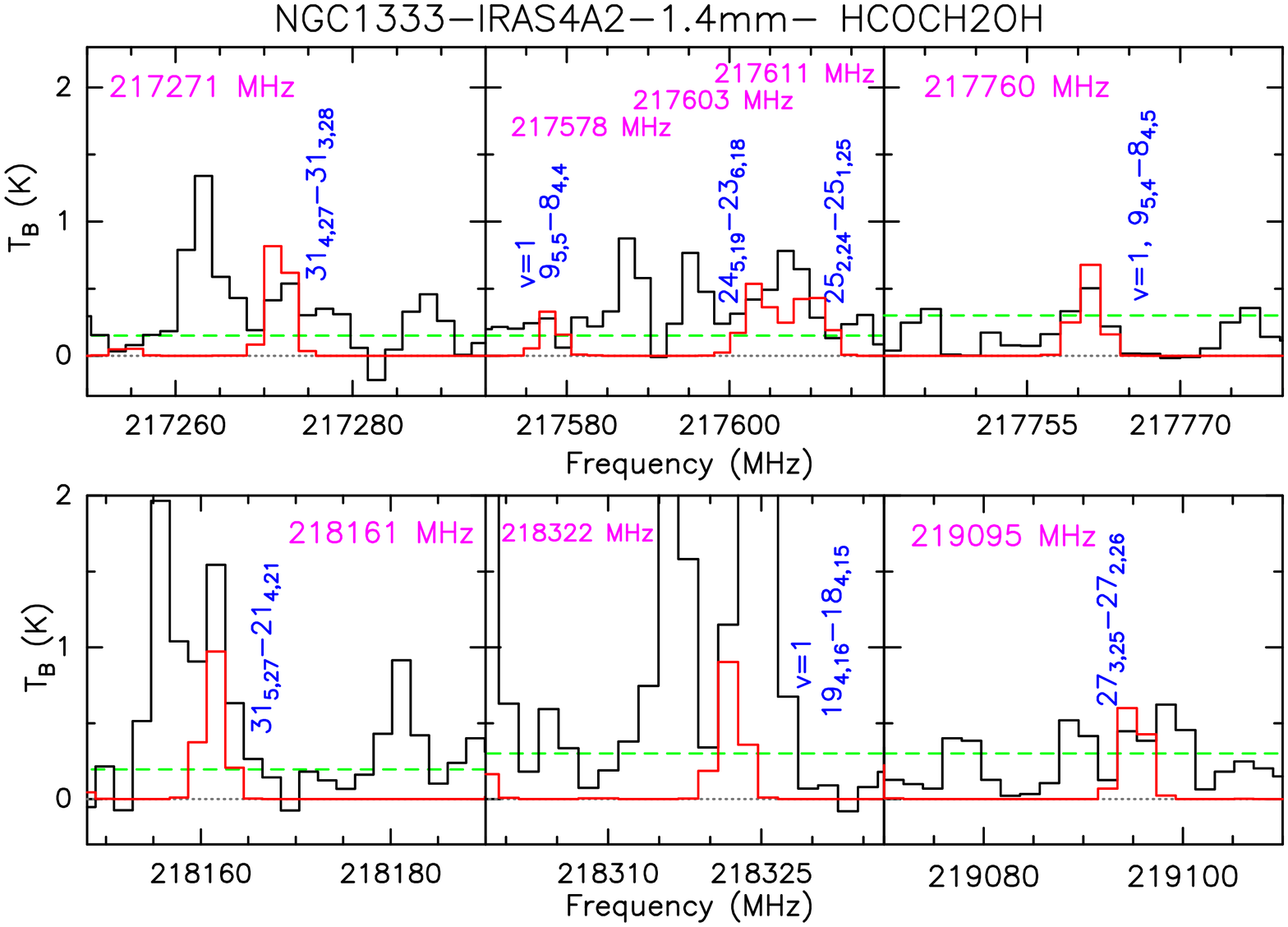}
\includegraphics[scale=0.48]{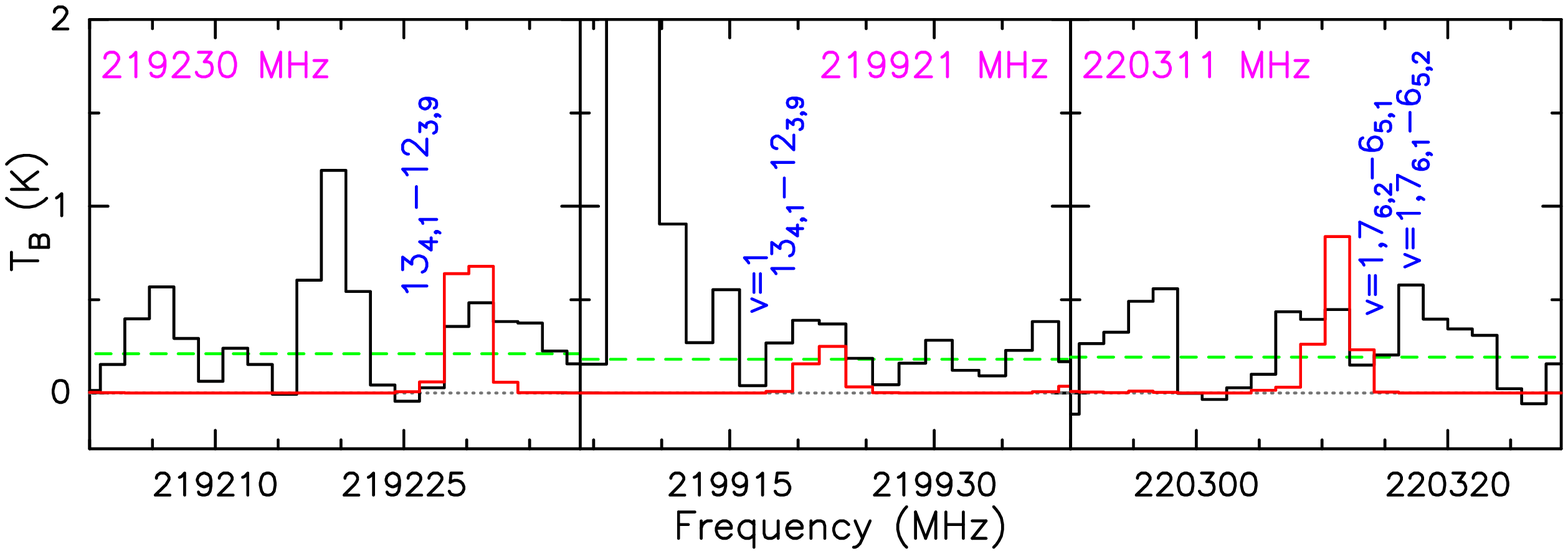}
\caption{Glycolaldehyde emission lines 
(in $T_{\rm B}$ scale)
observed in the 1.3mm and 1.4mm spectral windows towards 
NGC1333-IRAS4A2 and excluded from the LTE analysis because of 
severe blending with other emission lines. 
The horizontal green dotted lines show the 3$\sigma$ level.
In blue we mark the glycolaldehyde transitions.
The red line shows the synthetic spectrum obtained with the GILDAS--Weeds package 
(Maret et al. 2011) and assuming the rotation diagram solutions (see Fig. 3).}
\end{center}
\end{figure*}

\begin{figure*}
\begin{center}
\graphicspath{{figures/}}
\includegraphics[scale=0.52]{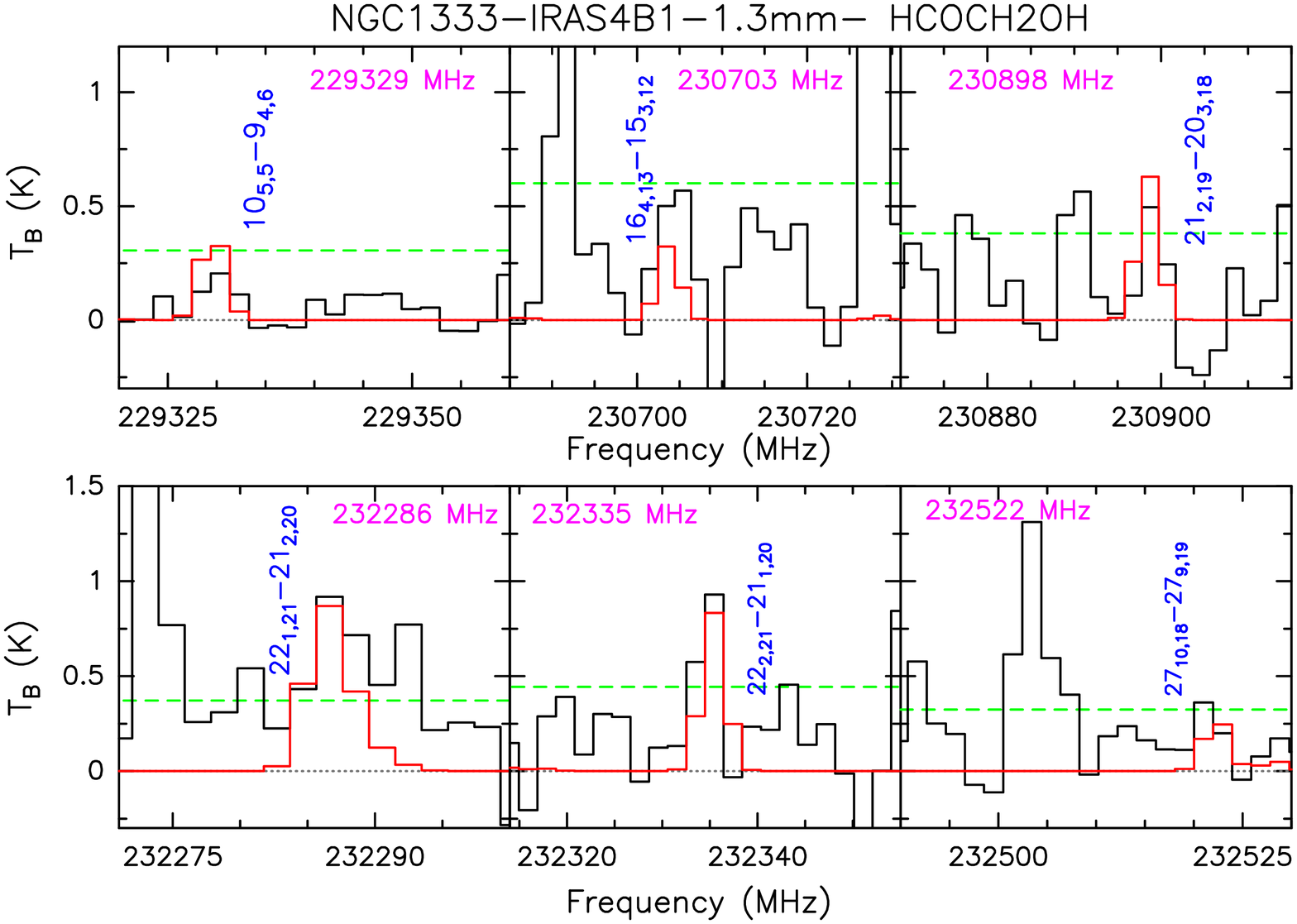} 
\includegraphics[scale=0.5]{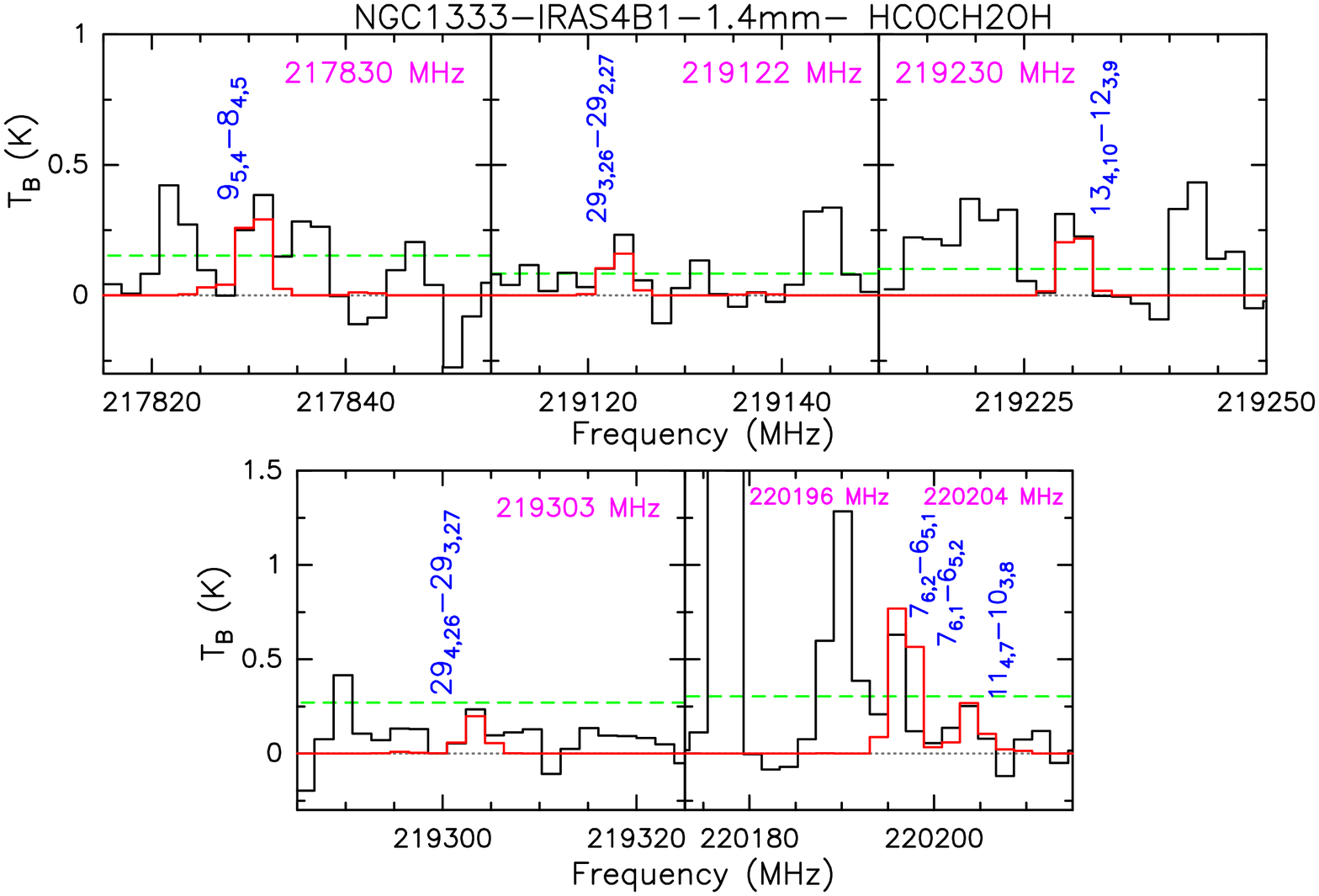}  
\caption{Glycolaldehyde emission lines (in $T_{\rm B}$ scale)
detected in the 1.3mm and 1.4mm spectral windows towards 
NGC1333-IRAS4B1 and used to perform the LTE analysis. 
The horizontal green dotted lines show the 3$\sigma$ level.
In blue we mark the glycolaldehyde lines extracted from Table A.3.
The red line shows the synthetic spectrum obtained with the GILDAS--Weeds package 
(Maret et al. 2011) and assuming the rotation diagram solutions (see Fig. 3).}
\end{center}
\end{figure*}

\begin{figure*}
\begin{center}
\graphicspath{{figures/}}
\includegraphics[scale=0.5]{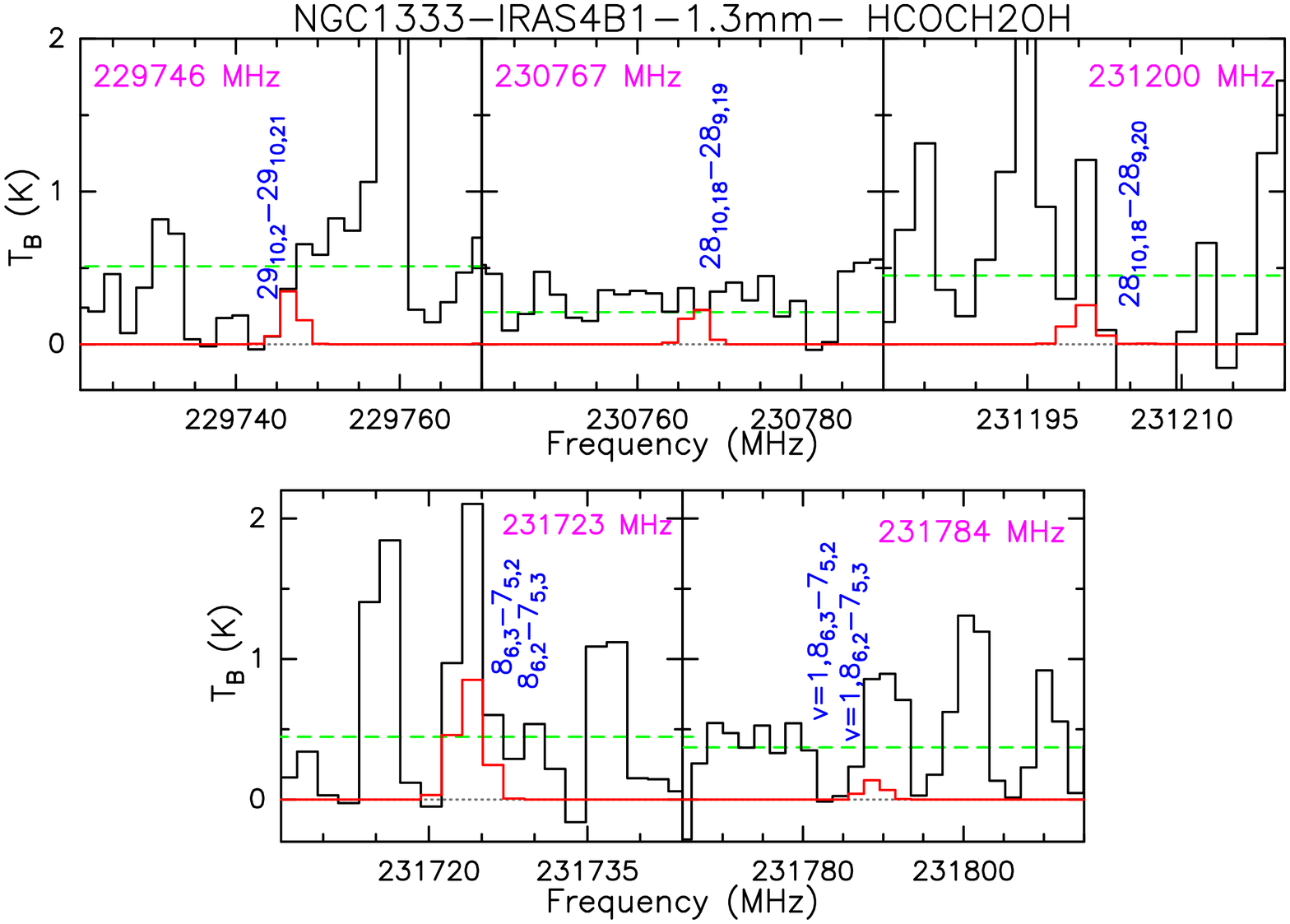}
\includegraphics[scale=0.5]{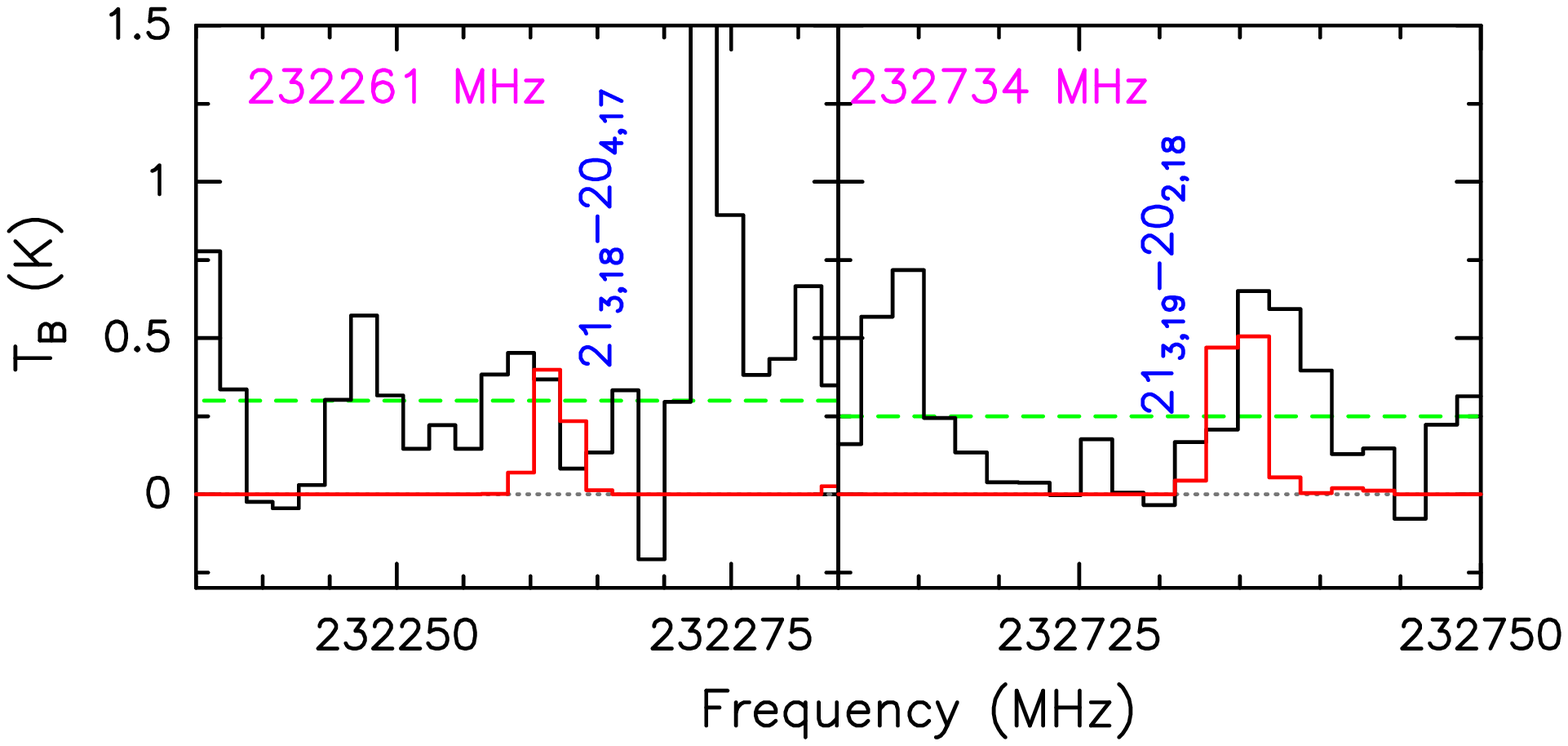}
\includegraphics[scale=0.5]{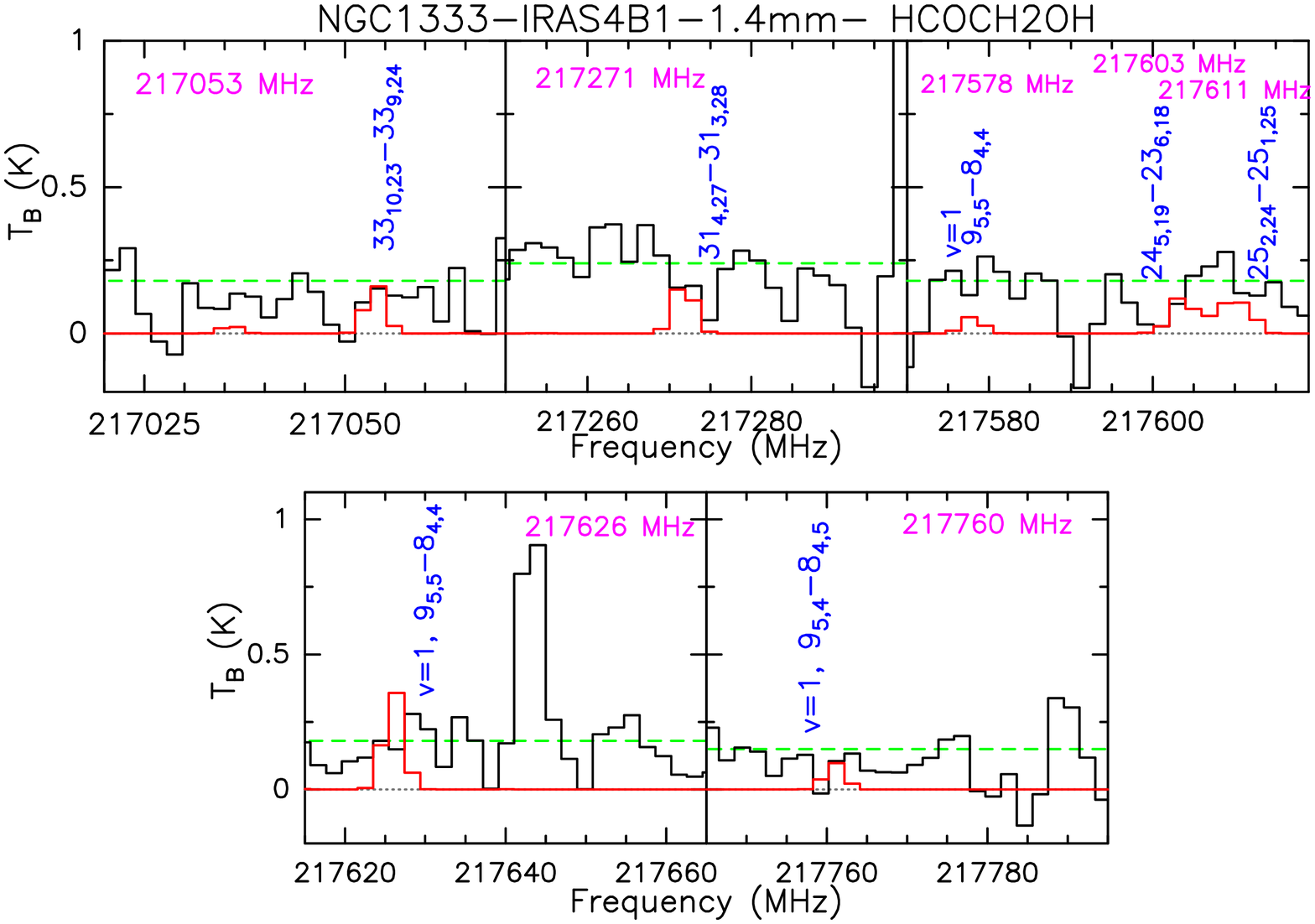}
\caption{Glycolaldehyde emission lines (in $T_{\rm B}$ scale)
observed in the 1.3mm and 1.4mm spectral windows towards 
NGC1333-IRAS4B1 and excluded from the LTE analysis because of 
severe blending with other emission lines. 
The horizontal green dotted lines show the 3$\sigma$ level.
In blue we mark the glycolaldehyde transitions.
The red line shows the synthetic spectrum obtained with the GILDAS--Weeds package 
(Maret et al. 2011) and assuming the rotation diagram solutions (see Fig. 3).}
\end{center}
\end{figure*}

\begin{figure*}
\addtocounter{figure}{-1}
\begin{center}
\graphicspath{{figures/}}
\includegraphics[scale=0.5]{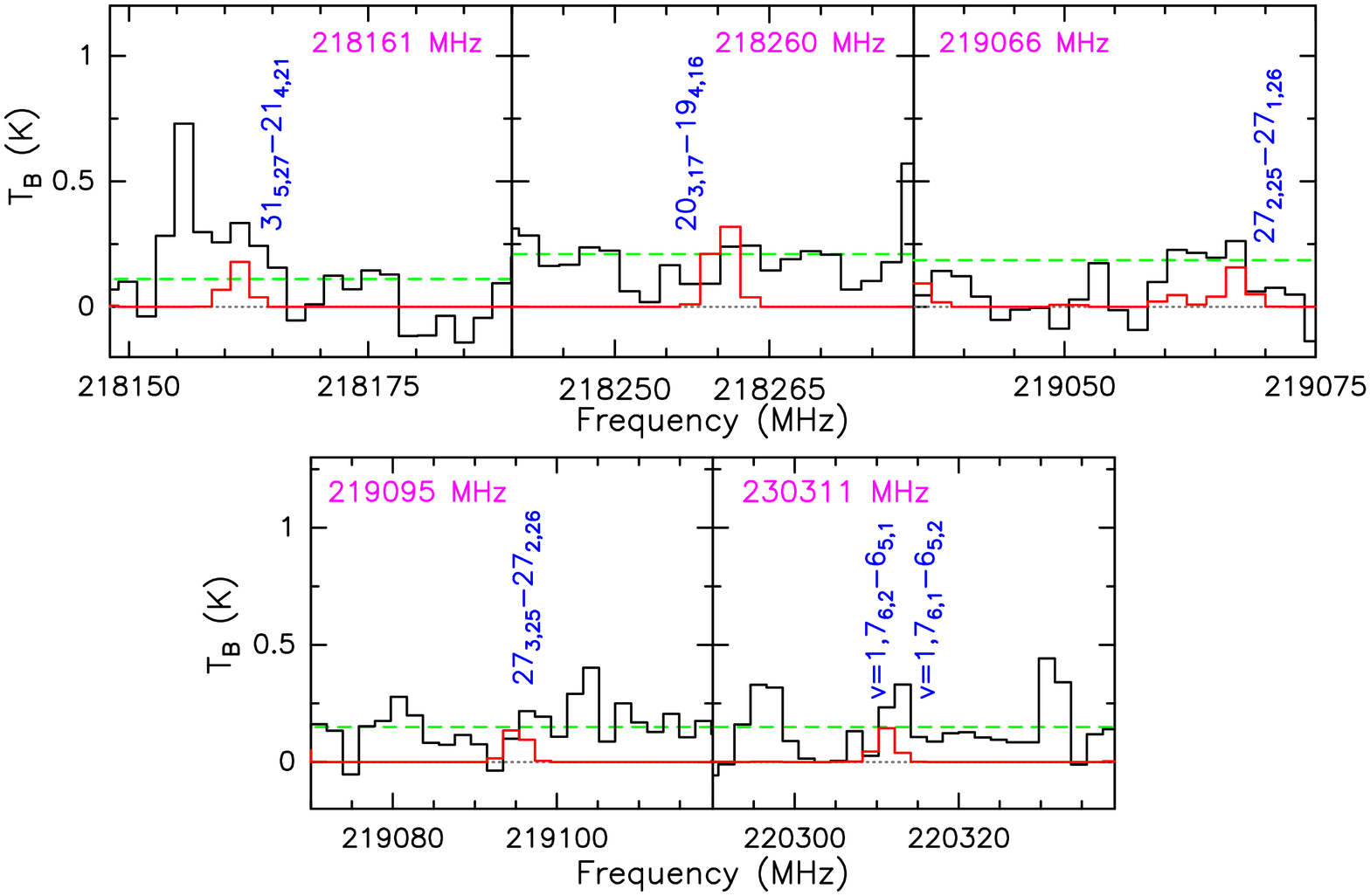}
\caption{{\it Continued}.}
\end{center}
\end{figure*}

\begin{figure*}
\begin{center}
\graphicspath{{figures/}}
\includegraphics[scale=0.5]{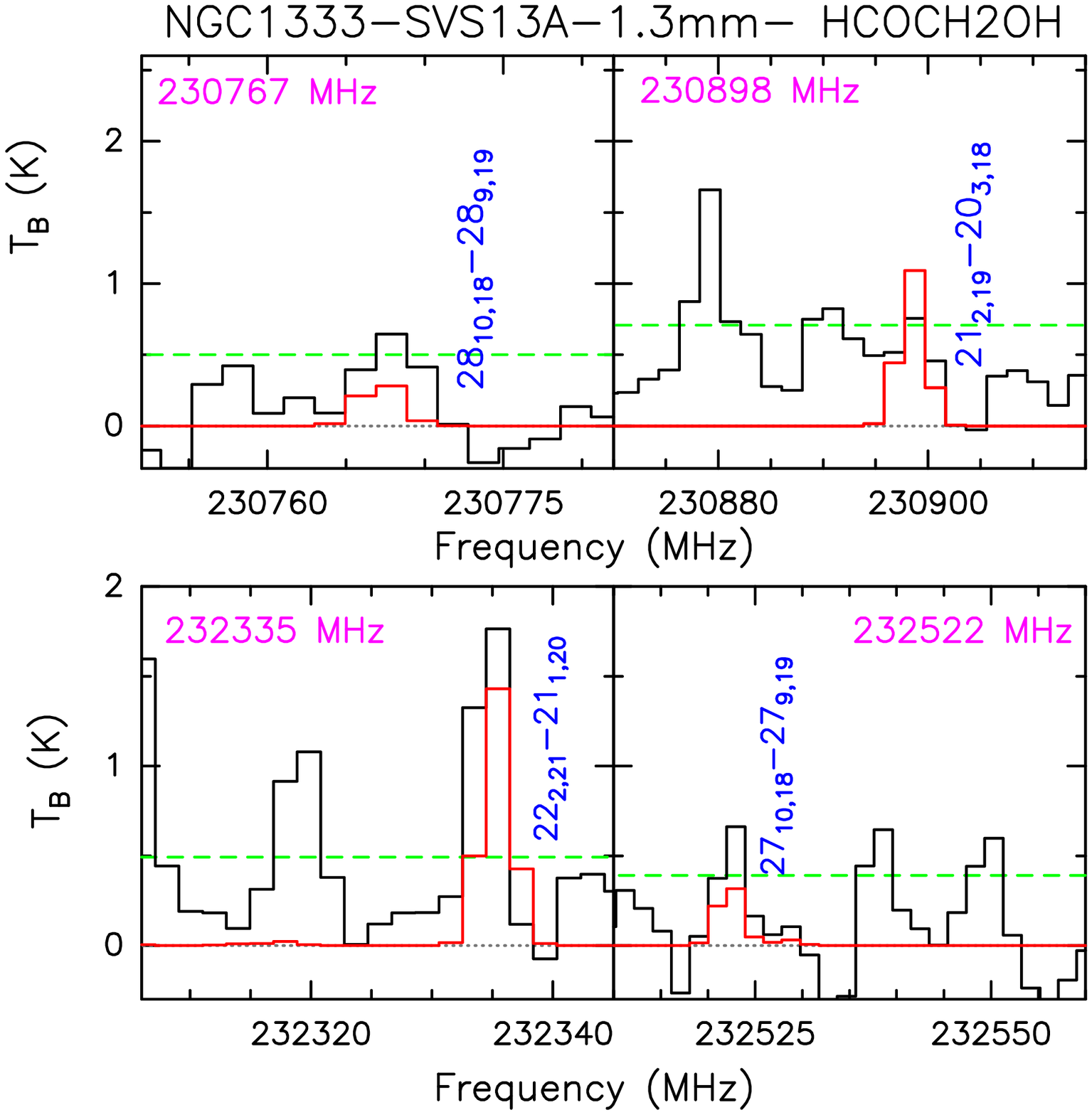} 
\includegraphics[scale=0.5]{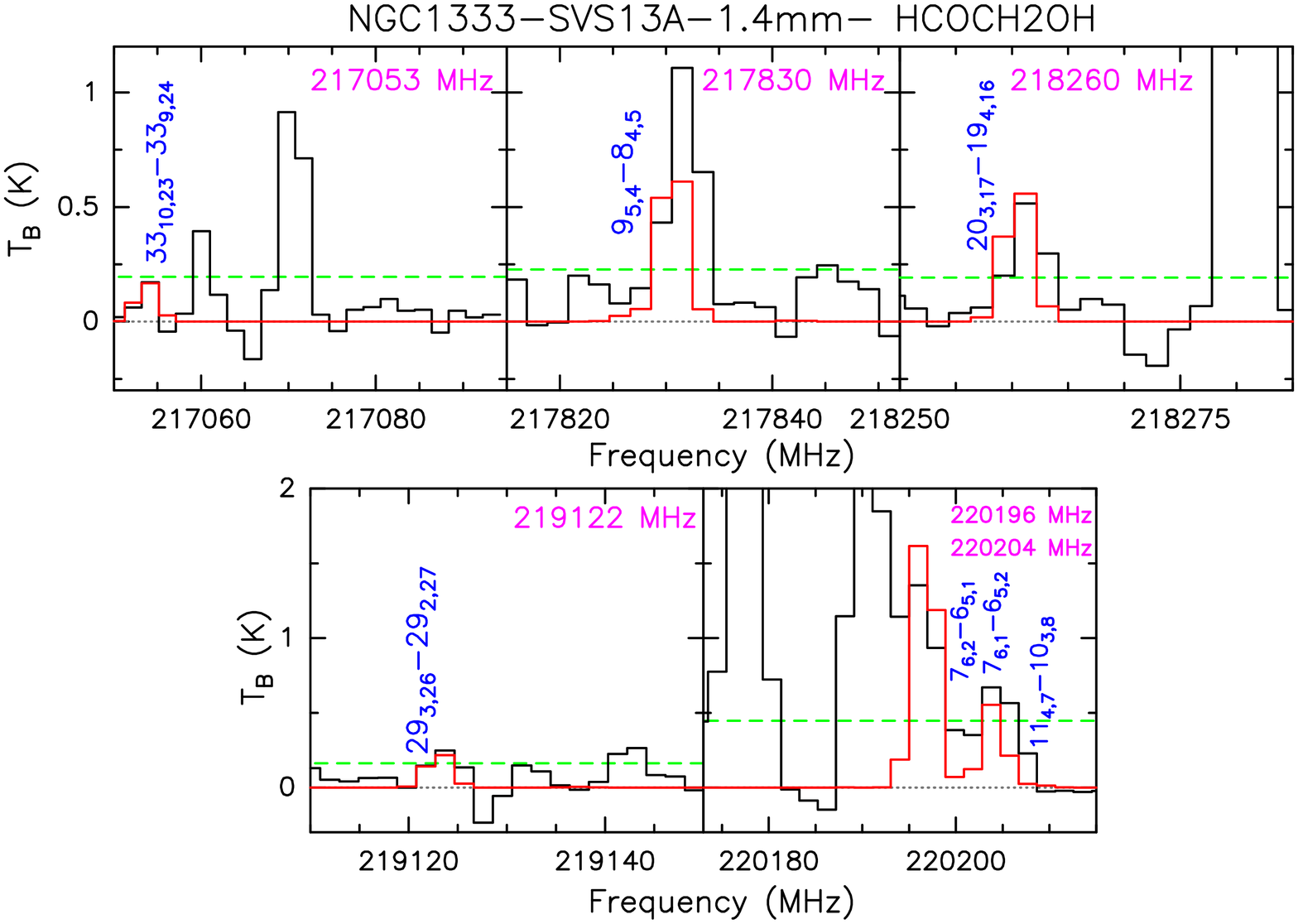}  
\caption{Glycolaldehyde emission lines (in $T_{\rm B}$ scale)
detected in the 1.3mm and 1.4mm spectral windows towards 
NGC1333-SVS13A and used  to perform the LTE analysis. 
The horizontal green dotted lines show the 3$\sigma$ level.
In blue we mark the glycolaldehyde lines extracted from Table A.4.
The red line shows the synthetic spectrum obtained with the GILDAS--Weeds package 
(Maret et al. 2011) and assuming the rotation diagram solutions (see Fig. 3).}
\end{center}
\end{figure*}

\begin{figure*}
\begin{center}
\graphicspath{{figures/}}
\includegraphics[scale=0.53]{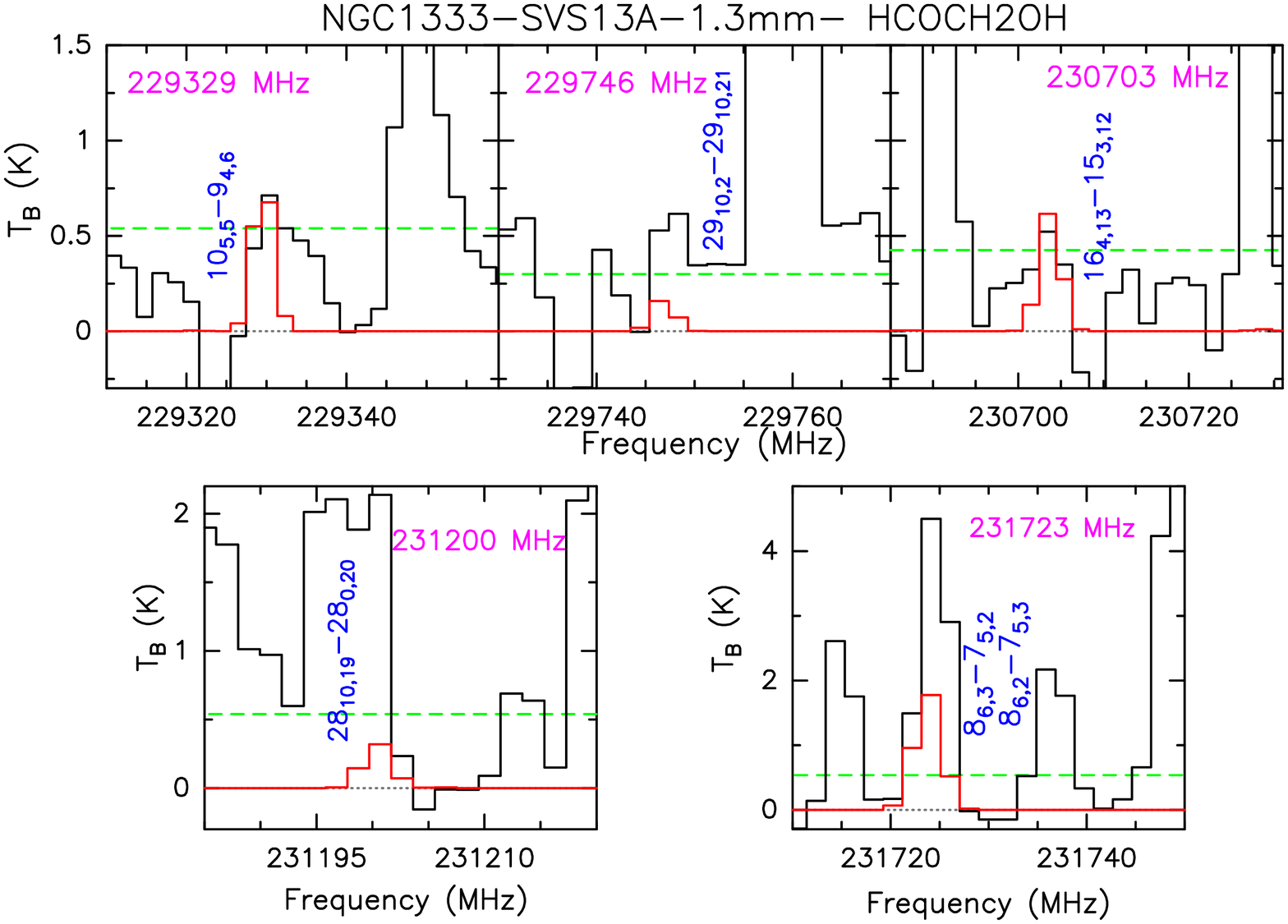}
\includegraphics[scale=0.53]{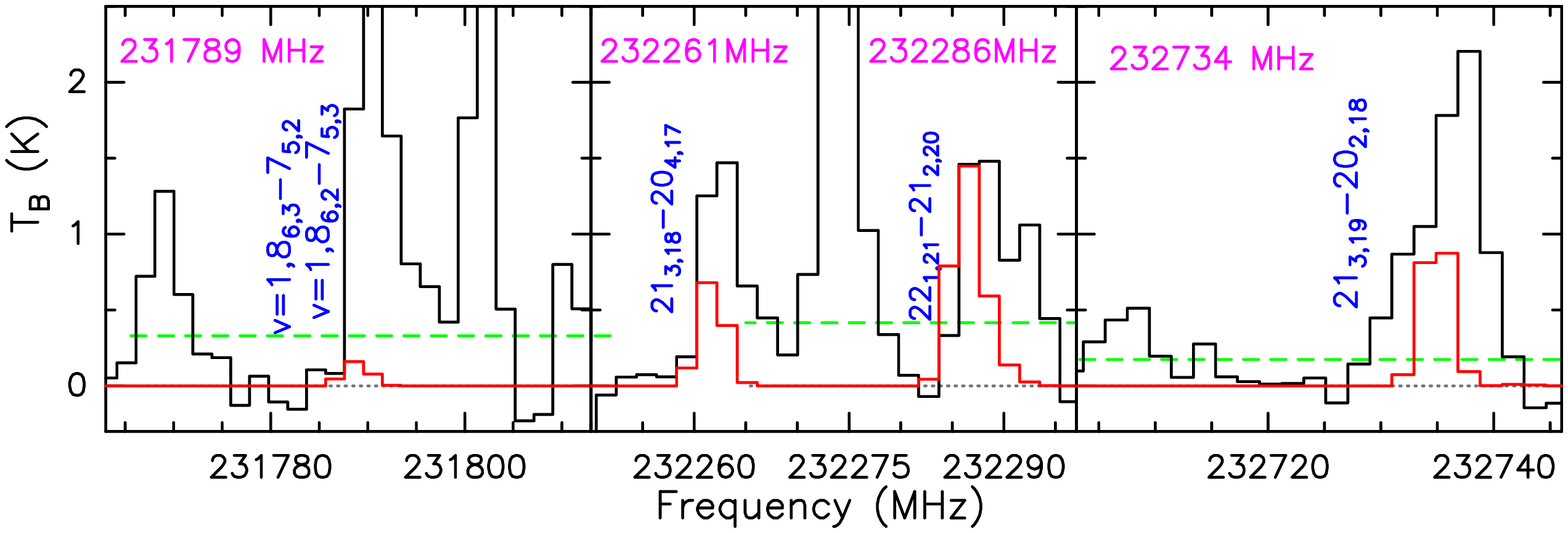}
\includegraphics[scale=0.53]{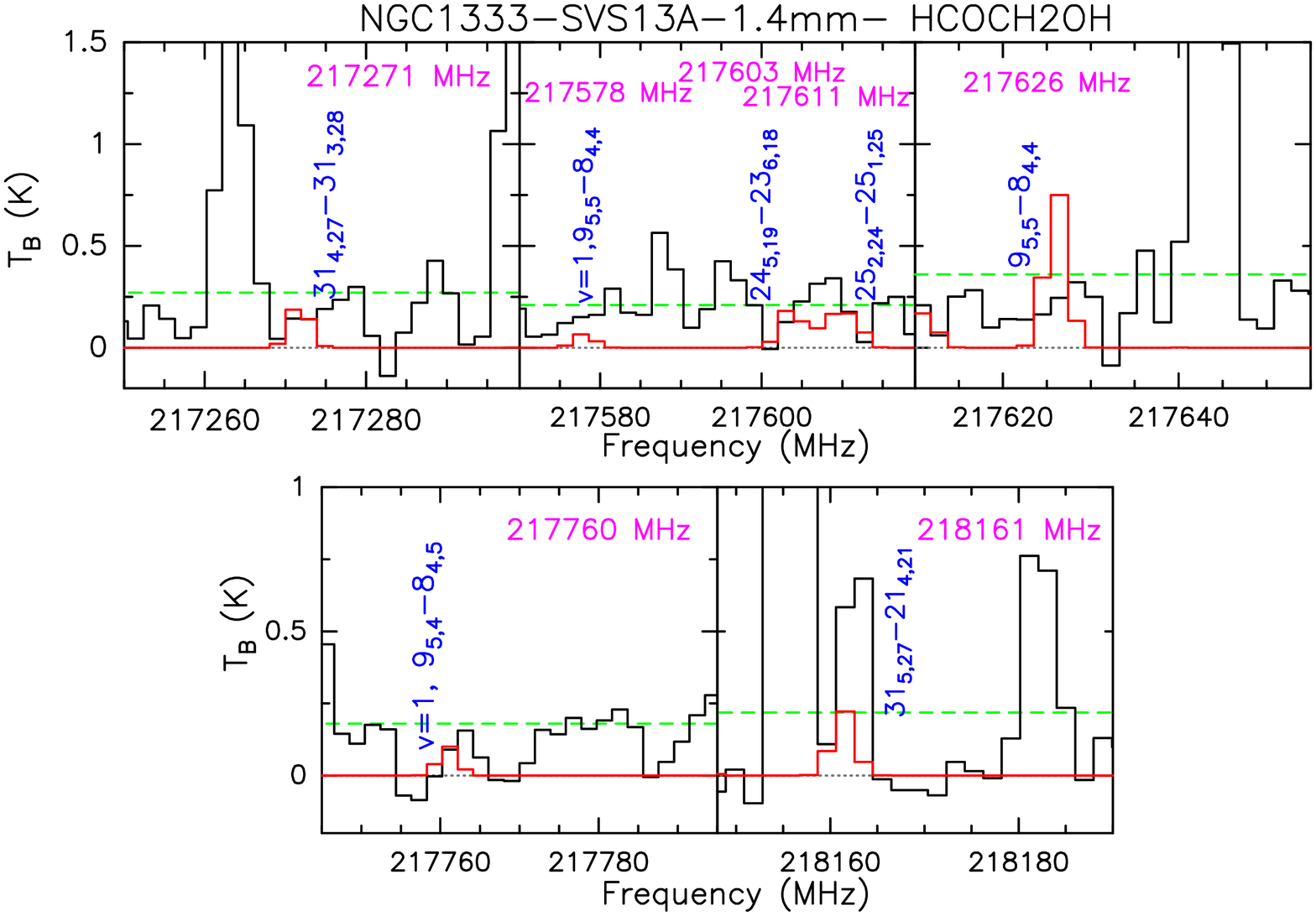}
\caption{Glycolaldehyde emission lines 
(in $T_{\rm B}$ scale)
observed in the 1.3mm and 1.4mm spectral windows towards 
NGC1333-SVS13A and excluded from the LTE analysis because of 
severe blending with other emission lines. 
The horizontal green dotted lines show the 3$\sigma$ level.
In blue we mark the glycolaldehyde transitions.
The red line shows the synthetic spectrum obtained with the GILDAS--Weeds package 
(Maret et al. 2011) and assuming the rotation diagram solutions (see Fig. 3).}
\end{center}
\end{figure*}

\begin{figure*}
\addtocounter{figure}{-1}
\begin{center}
\graphicspath{{figures/}}
\includegraphics[scale=0.53]{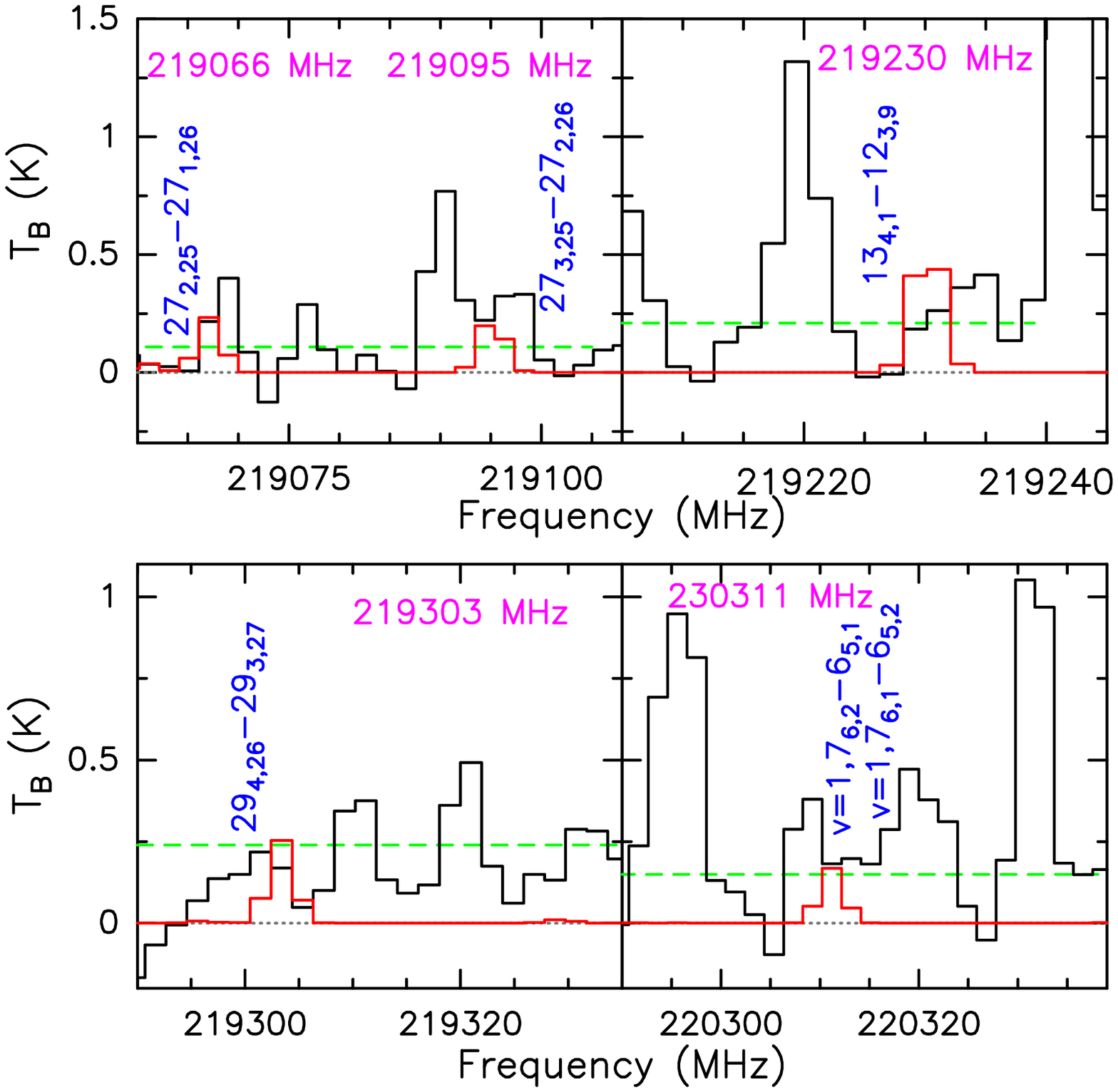}
\caption{{\it Continued}.}
\end{center}
\end{figure*}

\begin{figure*}
\begin{center}
\graphicspath{{figures/}}
\includegraphics[scale=0.5]{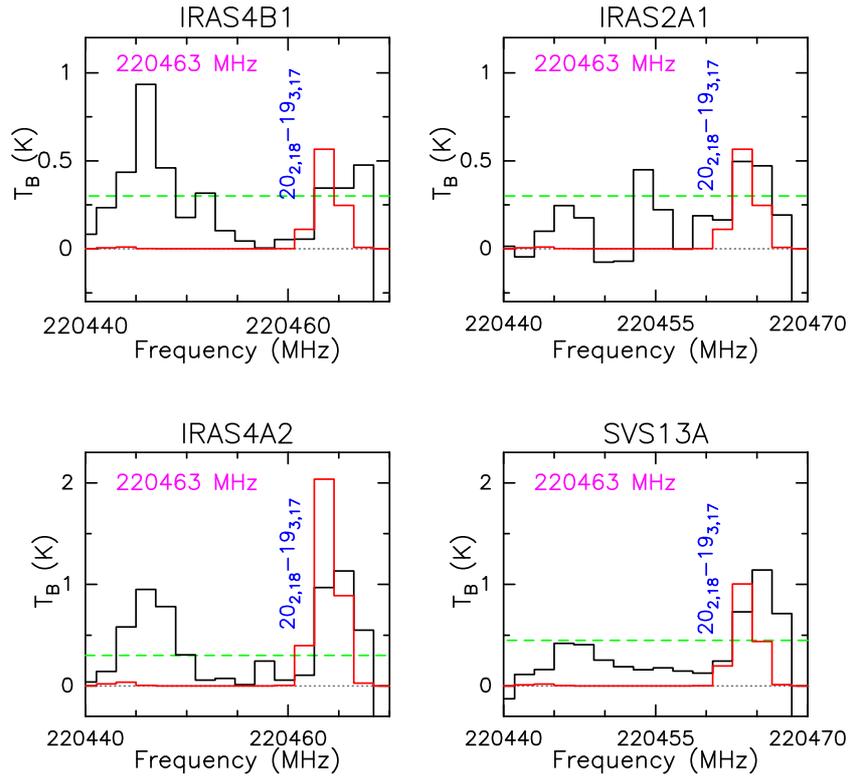}   
\caption{Glycolaldehyde 20$_{2,18}$-19$_{3,17}$ emission lines 
(in $T_{\rm B}$ scale)
observed at 220463 MHz  towards 
NGC1333-IRAS4B1, -IRAS2A1, IRAS4A2, and -SVS13A, and 
conservatively excluded from the LTE analysis falling at the edge of the observed WideX band.
The horizontal green dotted lines show the 3$\sigma$ level.
In blue we mark the glycolaldehyde transitions.
The red line shows the synthetic spectrum obtained with the GILDAS--Weeds package 
(Maret et al. 2011) and assuming the rotation diagram solutions (see Fig. 3).}
\end{center}
\end{figure*}

\clearpage

\begin{table*}
\caption{Glycolaldehyde emission lines detected towards NGC1333 IRAS2A1 (in $T_{\rm B}$ scale)}
\centering
\renewcommand\arraystretch{1.2}
\begin{tabular}{lcccccccc}
\hline
Transition$^a$ & $\nu$$^a$ & $E_{\rm u}$$^a$ & $S\mu^{2}$$^a$ & $\int{T dV}$$^b$ & $V_{\rm peak}$$^b$ & $FWHM$$^b$ & $T_{\rm peak}$$^b$ & rms \\
 & (MHz) & (K) & (D$^{2}$) & (K km s$^{-1}$) & (km s$^{-1}$) & (km s$^{-1}$) & (mK) & (mK) \\
\hline
 & & & & \hspace{-0.7cm}1.4mm & & & & \\
\hline
31$_{4,27}$-31$_{3,28}$  & 217271.520 & 290.40 & 63.50 & 1.8(1.0) & +6.5(1.2) & 4.1(2.9) & 434(32) & 125 \\
27$_{3,25}$-27$_{2,26}$ & 219095.117 & 207.92 & 31.71 & 2.3(0.6) & +7.4(0.6) & 4.9(1.6) & 438(32) & 104 \\
7$_{6,2}$-6$_{5,1}$  & 220196.605 & 37.41 & 29.67 & \multirow{2}{*}{10.5(3.8)} & \multirow{2}{*}{+7.8(0.9)} & \multirow{2}{*}{6.5(2.8)} & \multirow{2}{*}{1499(32)} & \multirow{2}{*}{100} \\
7$_{6,1}$-6$_{5,2}$  & 220196.800 & 37.41 & 29.67 \\
11$_{4,7}$-10$_{3,8}$  & 220204.039 & 46.61 & 18.68 & 4.3(2.8) & +8.0(1.9) &  7.0(6.4) & 578(31) & 100\\
7$_{6,2}$-6$_{5,1}$ $v$=1  & 220311.082 & 318.20 & 29.90 &  \multirow{2}{*}{2.7(1.4)} & \multirow{2}{*}{+6.9(1.6)} & \multirow{2}{*}{5.8(3.2)} & \multirow{2}{*}{422(34)} & \multirow{2}{*}{138} \\
7$_{6,1}$-6$_{5,2}$ $v$=1  & 220311.251 & 318.20 & 29.90 \\
\hline
 & & & & \hspace{-0.7cm}1.3mm & & & & \\
\hline
21$_{2,19}$-20$_{3,18}$  & 230898.536 & 131.19 & 71.15 & 4.2(2.2) & +6.0(1.6) & 5.9(3.4) & 670(80) & 167 \\
8$_{6,3}$-7$_{5,2}$  &  231723.266 & 41.85 & 29.86 & \multirow{2}{*}{16.6(4.7)} & \multirow{2}{*}{+7.2(1.2)} & \multirow{2}{*}{8.3(3.4)} & \multirow{2}{*}{1954(104)} & \multirow{2}{*}{220} \\
8$_{6,2}$-7$_{5,3}$  & 231724.332 & 41.85 & 29.86 \\
22$_{2,21}$-21$_{1,20}$  & 232335.395 & 134.52 & 94.55 & 8.9(2.5) & +7.8(0.6) & 4.6(1.6) & 1820(254) & 232\\ 
27$_{10,18}$-27$_{9,19}$  & 232522.255 & 271.37 & 77.57 & 2.9(1.9) & +8.4(2.0) & 6.1(4.8) & 454(78) & 107\\
21$_{3,19}$-20$_{2,18}$ & 232734.940 & 131.22 & 71.21 & 6.9(1.6)  & +6.2(0.6) & 4.8(1.1) & 1348(93) & 10 \\
\hline
\end{tabular}

$^a$ From the Jet
Propulsion Laboratory database (Pickett et al. 1998).
$^b$ The errors are the gaussian fit uncertainties. For the weakest
lines, the systemic velocity (see Table 1) and typical linewidths are assumed and
consequently no error on these parameters is quoted. 
\end{table*}

\begin{table*}
\caption{Glycolaldehyde emission lines detected towards NGC1333 SVS13-A (in $T_{\rm B}$ scale)}
\centering
\renewcommand\arraystretch{1.2}
\begin{tabular}{lcccccccc}
\hline
Transition$^a$ & $\nu$$^a$ & $E_{\rm u}$$^a$ & $S\mu^{2}$$^a$ & $\int{T dV}$$^b$ & $V_{\rm peak}$$^b$ & $FWHM$$^b$ & $T_{\rm peak}$$^b$ & rms \\
 & (MHz) & (K) & (D$^{2}$) & (K km s$^{-1}$) & (km s$^{-1}$) & (km s$^{-1}$) & (mK) & (mK) \\
\hline
 & & & & \hspace{-0.7cm}1.4mm & & & & \\
\hline
33$_{10,23}$-33$_{9,24}$  & 217053.668 & 374.45 & 104.58 & 0.6(0.2) & +8.4(0.0) &  3.5(0.0) & 184(21) & 65\\
9$_{5,4}$-8$_{4,5}$  & 217830.692 & 40.20 & 25.16 & 5.2(0.5) & +7.0(0.7) &  4.0(1.4) & 1096(36) & 76 \\
20$_{3,17}$-19$_{4,16}$  & 218260.540 & 126.14 & 43.31 &  2.3(0.7) & +7.1(0.7) &  4.5(1.4) & 455(70) & 64 \\
29$_{3,26}$-29$_{2,27}$  & 219122.861 & 247.80 &  47.54 &  1.3(0.5) & +7.4(1.0) &  4.7(1.5) & 270(18) & 54 \\
7$_{6,2}$-6$_{5,1}$  & 220196.605 & 37.41 & 29.67 & \multirow{2}{*}{5.6(1.8)} & \multirow{2}{*}{+8.5(0.0)} & \multirow{2}{*}{3.5(0.0)} & \multirow{2}{*}{1518(268)} & \multirow{2}{*}{149} \\
7$_{6,1}$-6$_{5,2}$  & 220196.800 & 37.41 & 29.67 \\
11$_{4,7}$-10$_{3,8}$  & 220204.039 & 46.61 & 18.68 & 2.5(1.7) & +8.4(0.0) &  3.5(0.0) & 685(268) & 149\\
\hline
 & & & & \hspace{-0.7cm}1.3mm & & & & \\
\hline
28$_{10,18}$-28$_{9,19}$  & 230767.152 & 287.09 & 81.84 & 3.9(1.2) & +7.3(0.9) & 5.7(1.9) & 630(221) & 167 \\  
21$_{2,19}$-20$_{3,18}$  & 230898.536 & 131.19 & 71.15 & 3.9(1.5) & +8.4(0.0) & 4.0(0.0) & 864(164) & 236 \\
22$_{2,21}$-21$_{1,20}$  & 232335.395 & 134.52 & 94.55 & 8.5 (1.8) & +9.3(0.3) & 3.8 (1.5) & 2097(45) & 164\\ 
27$_{10,18}$-27$_{9,19}$  & 232522.255 & 271.37 & 77.57 & 3.0(0.9) & +8.0(0.7) & 4.3(1.5) & 697(54) & 130\\
\hline
\end{tabular}

$^a$ From the Jet
Propulsion Laboratory database (Pickett et al. 1998).
$^b$ The errors are the gaussian fit uncertainties. For the weakest
lines, the systemic velocity (see Table 1) and typical linewidths are assumed and
consequently no error on these parameters is quoted. 
\end{table*}

\begin{table*}
\caption{Glycolaldehyde emission lines detected towards NGC1333 IRAS4A2 (in $T_{\rm B}$ scale)}
\centering
\renewcommand\arraystretch{1.2}
\begin{tabular}{lcccccccc}
\hline
Transition$^a$ & $\nu$$^a$ & $E_{\rm u}$$^a$ & $S\mu^{2}$$^a$ & $\int{T dV}$$^b$ & $V_{\rm peak}$$^b$ & $FWHM$$^b$ & $T_{\rm peak}$$^b$ & rms \\
 & (MHz) & (K) & (D$^{2}$) & (K km s$^{-1}$) & (km s$^{-1}$) & (km s$^{-1}$) & (mK) & (mK) \\
\hline
 & & & & \hspace{-0.7cm}1.4mm & & & & \\
\hline
33$_{10,23}$-33$_{9,24}$  & 217053.668 & 374.45 & 104.58 & 2.6(1.6) & +7.0(0.0) & 3.5(0.0) & 701(61) & 80\\
39$_{10,30}$-38$_{11,27}$  & 217626.061 & 40.20 & 25.19 & 4.2(1.1) & +5.9(0.6) & 4.5(1.3) & 876(20) & 90\\
9$_{5,4}$-8$_{4,5}$  & 217830.692 & 40.20 & 25.16 & 5.9(1.6) & +5.7(1.0) &  7.7(2.8) & 699(67) & 70 \\
20$_{3,17}$-19$_{4,16}$  & 218260.540 & 126.14 & 43.31 &  4.7(0.8) & +4.7(0.5) & 5.7(1.2) & 784(58)& 69 \\
27$_{2,25}$-27$_{1,26}$  & 219066.914 & 207.91 & 31.72 &  4.3(1.8) & +5.4(0.9) & 5.2(3.0) & 778(78) & 60\\
29$_{3,26}$-29$_{2,27}$  & 219122.861 & 247.80 &  47.54 &  2.4(1.2) & +6.2(1.2) & 5.4(3.0) & 404(63) & 49\\
29$_{4,26}$-29$_{3,27}$  & 219303.367 & 247.81 & 47.54 &  2.8(0.8) & +7.0(0.0) & 3.5(0.0) & 777(23) & 26 \\
7$_{6,2}$-6$_{5,1}$  & 220196.605 & 37.41 & 29.67 & \multirow{2}{*}{7.4(1.9)} & \multirow{2}{*}{+7.0(0.0)} & \multirow{2}{*}{3.5(0.0)} & \multirow{2}{*}{1969(256)} & \multirow{2}{*}{49} \\
7$_{6,1}$-6$_{5,2}$  & 220196.800 & 37.41 & 29.67 \\
11$_{4,7}$-10$_{3,8}$  & 220204.039 & 46.61 & 18.68 & 3.2(1.9) & +7.0(0.0) & 3.5(0.0) & 885(256) & 49\\
\hline
 & & & & \hspace{-0.7cm}1.3mm & & & & \\
\hline
10$_{5,5}$-9$_{4,6}$  & 229329.580 & 45.76 & 25.44 & 3.6(1.2) & +7.0(0.0) & 3.5(0.0) & 957(199) & 120  \\
16$_{4,13}$-15$_{3,12}$  & 230703.712 & 85.68 & 27.38 & 4.6(2.9) & +6.7(1.5) & 3.8(2.2) & 1102(197) & 180  \\
28$_{10,18}$-28$_{9,19}$  & 230767.152 & 287.09 & 81.84 & 3.1(1.2) & +7.0(0.0) & 3.5(0.0) & 866(149) & 200 \\  
21$_{2,19}$-20$_{3,18}$  & 230898.536 & 131.19 & 71.15 & 6.9(2.3) & +7.0(0.0) & 3.5(0.0) & 1887(133) & 181 \\
21$_{3,18}$-20$_{4,17}$  & 232261.725 & 137.84 & 49.86 & 8.6(3.6) & +6.1(0.9) & 3.9(2.0) & 2021(310) & 160 \\
22$_{1,21}$-21$_{2,20}$  & 232286.032 & 134.52 & 94.55 & 12.8(3.7) & +5.9(0.6) & 3.9(1.3) & 3031(31) & 150\\
22$_{2,21}$-21$_{1,20}$  & 232335.395 & 134.52 & 94.55 & 21.9(2.6) & +7.2(0.3) & 3.7(0.7) & 5592(23) & 211\\ 
27$_{10,18}$-27$_{9,19}$  & 232522.255 & 271.37 & 77.57 & 9.4(2.2) & +7.4(0.5) & 4.7(1.3) & 1861(62) & 124\\
\hline
\end{tabular}

$^a$ From the Jet
Propulsion Laboratory database (Pickett et al. 1998).
$^b$ The errors are the gaussian fit uncertainties. For the weakest
lines, the systemic velocity (see Table 1) and typical linewidths are assumed and
consequently no error on these parameters is quoted. 
\end{table*}

\begin{table*}
\caption{Glycolaldehyde emission lines detected towards NGC1333 IRAS4B1 (in $T_{\rm B}$ scale)}
\centering
\renewcommand\arraystretch{1.2}
\begin{tabular}{lcccccccc}
\hline
Transition$^a$ & $\nu$$^a$ & $E_{\rm u}$$^a$ & $S\mu^{2}$$^a$ & $\int{T dV}$$^b$ & $V_{\rm peak}$$^b$ & $FWHM$$^b$ & $T_{\rm peak}$$^b$ & rms \\
 & (MHz) & (K) & (D$^{2}$) & (K km s$^{-1}$) & (km s$^{-1}$) & (km s$^{-1}$) & (mK) & (mK) \\
\hline
 & & & & \hspace{-0.7cm}1.4mm & & & & \\
\hline
9$_{5,4}$-8$_{4,5}$  & 217830.692 & 40.20 & 25.16 & 1.8(0.5) & +7.0(0.0) &  3.5(0.0) & 485(23) & 51 \\
29$_{3,26}$-29$_{2,27}$  & 219122.861 & 247.80 &  47.54 &  1.0(0.5) & +6.2(1.2) &  4.2(1.9) & 226(7) & 28\\
13$_{4,10}$-12$_{3,9}$ & 219230.248 & 60.52 & 21.81 & 1.6(0.6) & +7.4(0.8) &  3.5(0.0) & 366(17) & 34 \\
29$_{4,26}$-29$_{3,27}$  & 219303.367 & 247.81 & 47.54 &  0.9(0.4) & +7.0(0.0) &  3.5(0.0) & 253(23) & 90 \\
7$_{6,2}$-6$_{5,1}$  & 220196.605 & 37.41 & 29.67 & \multirow{2}{*}{2.5(0.5)} & \multirow{2}{*}{+8.3(0.5)} & \multirow{2}{*}{3.7(0.8)} & \multirow{2}{*}{643(101) } & \multirow{2}{*}{101} \\
7$_{6,1}$-6$_{5,2}$  & 220196.800 & 37.41 & 29.67 \\
11$_{4,7}$-10$_{3,8}$  & 220204.039 & 46.61 & 18.68 & 1.3(0.6) & +7.8(1.0) &  4.6(2.2) & 251(101) & 101\\
\hline
 & & & & \hspace{-0.7cm}1.3mm & & & & \\
\hline
10$_{5,5}$-9$_{4,6}$  & 229329.580 & 45.76 & 25.44 & 3.7(0.9) & +6.2(0.8) & 5.2(1.6) & 638(65) & 102  \\
16$_{4,13}$-15$_{3,12}$  & 230703.712 & 85.68 & 27.38 &  3.4(1.8) & +6.1(1.4) & 5.2(2.8) & 631(56) & 206  \\
21$_{2,19}$-20$_{3,18}$  & 230898.536 & 131.19 & 71.15 & 2.2(0.9) & +6.2(0.9) & 3.7(1.5) & 551(25) & 127 \\
22$_{1,21}$-21$_{2,20}$  & 232286.032 & 134.52 & 94.55 & 2.9(0.9) & +7.0(0.0) & 3.5(0.0) & 699(124) & 124\\
22$_{2,21}$-21$_{1,20}$  & 232335.395 & 134.52 & 94.55 & 4.0(0.9) & +7.0(0.0) & 3.5(0.0) & 932(164) & 148\\ 
27$_{10,18}$-27$_{9,19}$  & 232522.255 & 271.37 & 77.57 & 1.8(0.9) & +8.3(1.4) & 4.5(2.8) & 387(108) & 108\\
\hline
\end{tabular}

$^a$ From the Jet
Propulsion Laboratory database (Pickett et al. 1998).
$^b$ The errors are the gaussian fit uncertainties. For the weakest
lines, the systemic velocity (see Table 1) and typical linewidths are assumed and
consequently no error on these parameters is quoted. 
\end{table*}

\end{document}